\documentclass[fleqn,usenatbib]{mnras}

\usepackage{newtxtext,newtxmath}
\usepackage{breqn}

\usepackage[T1]{fontenc}
\usepackage{ae,aecompl}

\usepackage{graphicx}	
\usepackage{amsmath}	
\usepackage{amssymb}	
\usepackage{float}

\title[Galaxy shape physics]{Prospects for Recovering Galaxy Intrinsic Shapes from Projected Quantities}

\author[R. Bassett \& C. Foster]{
Robert Bassett,$^{1,2}$\thanks{E-mail: rbassett@swin.edu.au}
Caroline Foster,$^{3,2}$
\\
$^{1}$Centre for Astrophysics and Supercomputing, Swinburne University
of Technology, PO Box 218, Hawthorn VIC 3122, Australia\\
$^{2}$ARC Centre of Excellence for All Sky Astrophysics in 3 Dimensions (ASTRO 3D)\\
$^{3}$Sydney Institute for Astronomy, School of Physics, A28, The University of Sydney, NSW, 2006, Australia
}

\pubyear{2018}

\begin{document}
\label{firstpage}
\pagerange{\pageref{firstpage}--\pageref{lastpage}}
\maketitle

\begin{abstract}
The distribution of three dimensional intrinsic galaxy shapes has been a longstanding open question. The difficulty stems from projection effects meaning one must rely on statistical methods applied to galaxy samples to infer intrinsic shape distributions. Theoretical work using analytical galaxy potentials suggests a relationship between galaxy intrinsic shape (as defined by its ``triaxiality'', in practice a proxy for how prolate a galaxy is) and the intrinsic misalignment angle between kinematic and morphological axes ($\Psi_{\rm int}$). This relationship reduces the number of unknowns, providing more reliable inferred intrinsic shape distributions than methods using photometry alone. Here we explore the connection between intrinsic shape and stellar kinematics using cosmological hydrodynamical simulations from the Illustris project.
The strongest relationship we find is that galaxy intrinsic flattening is correlated with specific angular momentum ($j$) with high $j$ galaxies being flatter than galaxies with low specific angular momentum. 
Our analysis shows that, although the majority of kinematically misaligned galaxies exhibit prolate shapes, examples of kinematically aligned prolate galaxies are also present. Clearly a direct correspondence between prolate shape and minor-axis rotation (often referred to as ``prolate rotation") is not present in Illustris. Thus, we demonstrate that the assumption of a simple relationship between $\Psi_{\rm int}$ and intrinsic shape commonly employed in shape recovery studies is not valid for Illustris galaxies. 
We suggest improvements on the method as well as some alternative methods for future work in this area.
\end{abstract}




\section{Introduction}


Information about the evolution of a galaxy can be encoded in its 3D shape. Galaxy shape has been shown to relate closely to a variety of galaxy properties including age \citep{vandeSande18}, spin \citep[e.g.][]{Weijmans14,Foster16,Foster17,Li18b}, stellar mass \citep[e.g.][]{SanchezJanssen10, Holden12, Ene18}, luminosity \citep[e.g.][]{SanchezJanssen16}, morphology \citep[e.g.][]{Ryden06,Padilla08,Rodriguez13}, and environment \citep[e.g.][]{Ryden93,Fasano10,Rodriguez16}. Furthermore, simulations suggest shape depends on merger history, in particular the mass ratio and configuration of the most recent galaxy merger \citep{Jesseit09,Taranu13,Moody14,Li18a,Li18b}. 


Early attempts of galaxy intrinsic shape recovery focused on comparing the distribution of apparent axis ratios for observed samples with the expected apparent axis ratio distributions for randomly oriented ellipsoids where the intrinsic axis ratios can vary.
This method was employed assuming that galaxies are either rotationally symmetric \citep[i.e. prolate or oblate,][]{Sandage70,Binney78} or triaxial \citep[e.g.][among others]{Benacchio80,Binney81,Fasano91,Ryden96,Vincent05,Kimm07,Padilla08}. 

Later methods incorporating galaxy kinematics were also developed that have the advantage of eliminating most of the unknowns through marginalisation \citep{Binney85,Franx91,Statler94a,Statler94b,Statler94c}. Kinematic shape recovery methods rely on accurate characterization of the apparent rotational axes of galaxies. Until recently this was only possible using radio interferometric observations of very nearby galaxies \citep[e.g.][]{Bak00}, however, with the advent of integral field spectroscopic (IFS) surveys \citep{Cappellari11,Croom12,Ma14} such measurements for large galaxy samples are now available. Recent IFS shape recovery studies such as \citet{Weijmans14}, \citet{Foster17}, \citet{Li18a} and \citet{Ene18} represent the current state-of-the-art, though these techniques have significant caveats. For a review of the history of galaxy shape recovery studies, see \citet{MendezAbreu16}.


From a theoretical perspective, many works have explored the shapes of galaxies in hydrodynamical simulations \citep[e.g.][]{Naab03,Jesseit09,Naab14,Ebrova15,Schaller15,Li16,Ebrova17,Li18b}. Together, these results inform our understanding of the connection between galaxy 3D shape and kinematics on which IFS shape recovery methods rely. The key analysis missing, however, is a direct test of IFS shape recovery methods using large samples of randomly-oriented, mock IFS observations designed to mimic observations from current and future IFS galaxy survey \citep[in a similar vein to the analysis of galaxy spin parameter presented by][]{Harborne18}. 


In this paper we explore the 3D versus 2D (projected) shapes of galaxies from the Illustris cosmological hydrodynamical simulations \citep{Genel14,Vogelsberger14a,Vogelsberger14b} in the context of shape recovery studies employing IFS observations. In particular we test the validity of the assumed relation between galaxy ``triaxility'', $T$, and kinematic misalignment, $\Psi_{\rm int}$ (see Sections \ref{section:3dmeasure} and \ref{section:2p2} for definitions), which is integral to the recovered intrinsic shapes in recent IFS studies \citep{Weijmans14,Foster17,Ene18}. This relation, based on theoretical work of \citet{Franx91}, implicitly includes two possibly problematic features: first, small deviation from a perfectly round face-on projection requires a large change in $\Psi_{\rm int}$ and second, prolate galaxies (which exhibit the largest $T$) must have $\Psi_{\rm int} \simeq 90^{\circ}$. We find that, as a class, prolate galaxies are of particular interest given this latter requirement, thus we also explore the origins of different $\Psi_{\rm int}$ values found for prolate galaxies.

After checking the validity of the assumed kinematics-shape connection, we then test the shape recovery method using projected stellar luminosity and kinematics maps produced from Illustris halo particle data. We show that employing the kinematic-shape relation during intrinsic shape recovery introduces strict biases in line with the problematic features highlighted above. We then test whether removing the assumption \citep[similar to the work of][]{Li18a} provides a slightly more reliable recovery of galaxy shape. 

This paper is organised as follows: In Section \ref{section:method} we describe the simulations and the subset of Illustris galaxies used as well as the methodology of our analysis, in Section \ref{section:results} we present the results of our 3D shape and kinematic analysis, in Section \ref{section:altass} we present the results of our shape recovery method from 2D mock observables, in Section \ref{section:discussion} we provide additional discussion of our results with a particular focus on the comparison between shapes in Illustris and shapes of galaxies in observations, and in Section \ref{section:conclusions} we summarise our results and enumerate our conclusions.

\section{Simulations and method}\label{section:method}

Our simulated galaxy data comes from the Illustris project \citep{Vogelsberger14a,Vogelsberger14b,Genel14}, which comprises a suite of hydrodynamic galaxy formation simulations in a cosmological volume using the moving mesh code, AREPO \citep{Springel10}. Galaxy samples produced in Illustris are found to be well matched to a variety of observations including the mass-size relation, galaxy luminosity function, and cosmic star-formation rate density \citep[among others][]{Vogelsberger14b,Snyder15,Xu17}. This work employs the largest Illustris simulation, Illustris 1, the details of which can be found in \citet{Vogelsberger14a} and data are taken from the Illustris public data release \citep{Nelson15}. 

We select galaxies from the final snapshot of Illustris 1, snapshot 135, corresponding to the local, $z=0$ universe. We then select only those galaxies with $\geq10^{5}$ stellar particles in order to ensure that measured galaxy shapes, both in two and three dimensions (see Sections \ref{section:3dmeasure} and \ref{section:2dmeasure}), are robust. This lower limit on particle number roughly corresponds to a stellar mass limit of our sample of $\sim10^{10}$ $M_{\odot}$ similar to cuts used in other works exploring galaxy shape in Illustris \citep[e.g.][]{Li18b}. This results in a sample of 978 galaxies for our analysis.

\subsection{Measuring 3D Shape}\label{section:3dmeasure}

We assume that the stellar content of galaxies can be approximated as simple ellipsoids with three principal axes of lengths $a \ge b \ge c$. In doing so, we ignore the well known fact that galaxies can contain multiple stellar components (e.g. bulge and disc) and focus on the ellipsoid-equivalent shape. With this simplifying assumption, the intrinsic shape of a galaxy can be fully parametrised with the two intrinsic axis ratios $p=b/a$ and $q=c/a$, such that $0\le q\le p\le1$.

We determine the ellipsoidal axis ratios for stellar particles in our sample of Illustris galaxies using the reduced inertia tensor \citep[similar to][]{Allgood06,Li18b}. The reduced inertia tensor, $I_{ij}$, is defined as:
\begin{equation}\label{eq:I}
	I_{ij} \equiv \sum_{n} \frac{x_{i,n}x_{j,n}}{\tilde{r}_{n}}
\end{equation}
where $\tilde{r}_{n}$ is the 3D, ellipsoidal radius to stellar particle $n$,
\begin{equation}
\tilde{r}_{n}=\sqrt{x_{n}^{2}+(y_{n}/p)^{2}+(z_{n}/q)^{2}} 
\end{equation}
where the $x$, $y$, and $z$ directions are aligned with the major, intermediate, and minor ellipsoidal axes. The eigenvectors of $I_{ij}$ associated with the lowest and highest eigenvalue represent the minor and major axes of the best fitting ellipsoid, respectively. The axis ratios $p$ and $q$ are computed as the square root of the ratios of the corresponding eigenvalues. 

For galaxies produced in hydrodynamic simulations, reliable measurements of $p$ and $q$ require a careful exclusion of stellar particles at large radii where asymmetries can have a large impact on $I_{ij}$. We do this by selecting only those particles within the ellipsoidal half mass radius, $\tilde{r}_{e}$.
The galaxy shape is then measured from the eigenvector of $I_{ij}$ computed using only selected stellar particles.

In practice we must employ an iterative process beginning with an initial guess of $a = b = c$ (i.e. a sphere), as these inputs for selecting the ellipsoidal half mass radius are unknown to start. By preselecting stellar particles in a spherical aperture, we bias this initial measurement of $p$ and $q$ towards a spherical ellipsoid. After fitting for the axes of the stellar particles we rotate the ensemble such that the major, intermediate, and minor axes are aligned with the $x$, $y$, and $z$ directions, respectively. We then recompute the ellipsoidal half mass radius using updated values of $p$ and $q$ from the previous iteration to evaluate $\tilde{r}_{e}$ and determine the eigenvalues of $I_{ij}$ using the new selection of stellar particles. This process is repeated until the values of $p$ and $q$ converge.



\subsection{Measuring $\Psi_{\rm int}$}\label{section:2p2}


The intrinsic kinematic misalignment of a galaxy, $\Psi_{\rm int}$, is defined as the angle between the short axis of the equivalent ellipsoid (see Section \ref{section:3dmeasure}) and the stellar ``angular rotation vector'', $\vec{R}$, in 3D. We note that $\vec{R}$ is measured only for those stellar particles within $\tilde{r}_{e}$ (i.e. the same particles used to measure $p$ and $q$ above). $\vec{R}$ is defined as:
\begin{equation}
	\vec{R} \equiv \sum_{n} \vec{r_{n}} \times L_{n}\vec{v_{n}}
\end{equation}
and is representative of the angular rotation of all stellar particles within an ellipsoidal half mass radius for a given galaxy. Here $L_{n}$ is the $r$-band luminosity of stellar particle $n$. 

$\vec{R}$ is analogous to the angular momentum with stellar mass replaced by $r$-band luminosity. This is done to provide the most reasonable comparison to mock observations from which the 2D projected $\Psi$ (see Section \ref{section:2dmeasure}) is measured from a stellar luminosity-weighted, projected kinematic map. We also note that although $\vec{R}$ is calculated using spherical radii rather than ellipsoidal (like $I_{ij}$), using ellipsoidal radii will not have an appreciable impact on the direction of $\vec{R}$. As we are concerned only with the direction of $\vec{R}$, this subtlety will have no effect on our results. We then define $\Psi_{\rm int}$ as:
\begin{equation}\label{eq:3dpsiint}
	\Psi_{\rm int} = 90^{\circ} - \cos^{-1}\left(\frac{\vec{R}\cdot\vec{e_{1}}}{|\vec{R}|}\right)
\end{equation}
where $\vec{e_{1}}$ is the unit vector parallel to the major axis of the ellipsoid fit following the procedure in Section \ref{section:3dmeasure}. The leading $90^{\circ}$ on the left of Equation \ref{eq:3dpsiint} is necessary such that $\Psi_{\rm int}$ is defined as $0^{\circ}$ when rotation is about the minor axis and $90^{\circ}$ when rotation is about the major axis. 

We define $\Psi_{\rm int}$ based on the major axis rather than the minor axis due to the fact that for near axisymmetric prolate galaxies ($p=q<1$), the minor axis direction is not well defined. For kinematically aligned, axisymmetric, prolate galaxies, where one expects $\vec{R}$ to align with the minor axis, defining $\Psi_{\rm int}$ based on the poorly defined minor axis direction may result in an incorrect identification as kinematically misaligned. We show in Section \ref{section:results} that the majority of kinematically misaligned galaxies are, in fact, prolate, which motivates our definition of $\Psi_{\rm int}$ based on the major axis in order to avoid incorrect kinematic classifications. A similar issue could arise when defining $\Psi_{\rm int}$ using the major axis for axisymmetric, oblate ($q<p$, $p=1$), kinematically misaligned galaxies, however, we  show in Section \ref{section:results} that oblate galaxies are almost exclusively kinematically aligned (particularly where $p\simeq 1$). Thus, incorrect kinematic classifications for oblate galaxies do not pose a major problem.


\subsection{Mock observations}\label{section:2dmeasure}

Here we describe first our method of producing projected 2D stellar luminosity and kinematics maps. From these maps we then measure kinematic offsets similar to those observed in IFS surveys \citep[e.g.][]{Weijmans14,Foster16,Foster17,Ene18}

\subsubsection{Stellar Luminosity and Kinematics Maps}\label{section:mocks}


We produce mock observables from Illustris stellar particle data in order to more directly compare with methods and selections employed in observational studies of galaxy shape. Specifically, we produce stellar luminosity, stellar velocity, and stellar velocity dispersion, $\sigma_{*}$, maps for each galaxy in our study.

The Sydney Australian-Astronomical-Observatory Multi-object IFS (SAMI) Galaxy Survey \citep{Croom12, Green18, Scott18} is a recently completed IFS survey of $\sim 3000$ nearby ($0.004 < z < 0.113$) galaxies. The mock observables we produce herein are designed to roughly match data products of the SAMI survey as galaxies from SAMI are the focus of a recent kinematic shape recovery effort \citep{Foster17}. SAMI data have a spaxel scale of 0$\farcs$5, mean photometric seeing of 2$\farcs$06 and a median redshift $z\simeq0.05$ \citep{Scott18}. At this redshift, the spaxel scale and average seeing correspond to  0.492 kpc and 2.03 kpc, respectively. We adopt these spatial scales for producing luminosity and kinematics maps of our Illustris galaxy sample.

To create a stellar luminosity map we first subtract the median x, y, and z positions and velocities from each star particle to center the galaxy spatially and shift it to its systemic velocity. We then produce a 2D grid of pixels 24.108 kpc on a side with a grid spacing of 0.492 kpc (49x49 pixels) on which we sample star particle positions. Then, at each pixel we sum the $r$-band luminosity of star particles with x and y locations placing them inside that pixel. To simulate the 2$\farcs$06 seeing typical of SAMI observations, we convolve the summed luminosity grid with a 2D Gaussian profile with a full width half max (FWHM) of 2.4$\times\sigma$.

Stellar kinematic maps are produced on the same 2D grid used for the luminosity maps with each spaxel position representing a line-of-sight velocity distribution (LOSVD) of star particles around that spatial location. We match the sampling of LOSVD's for each spaxel to the spectral sampling of SAMI datacubes of 0.57 {\AA} per spectral pixel, or $\sim$8.8 km/s. To produce kinematic maps we iterate through each pixel of our stellar luminosity maps and
create a 2D Gaussian at the pixel location with a $\sigma$ equivalent to the average seeing. We then weight each stellar particle in the galaxy by the value of this Gaussian at its projected 2D position multiplied by its $r$-band luminosity. Next we produce a weighted histogram of the line-of-sight (z-component) velocity of all particles in the galaxy with a bin size of 8.8 km/s. In this way we create a LOSVD at each spatial pixel with the effects of seeing included.

We then fit each LOSVD using a Gaussian profile with the central velocity and Gaussian $\sigma_*$ giving the stellar velocity and velocity dispersion at each position. Here we employ a fixed artificial detection limit to our data that is roughly matched the observations from the SAMI survey and corresponds to a mass surface density limit of 2$\times$10$^{8}$ $M_{\odot}$ arcsec$^{-2}$. We add Gaussian noise to each LOSVD corresponding to the average peak of the LOSVD in locations having this detection limit then subsequently fit each LOSVD with a single Gaussian. From these fits we extract the central velocity and velocity dispersion of each simulated spaxel, thus producing stellar velocity and velocity dispersion maps of each projection. 

\begin{figure*}
	\includegraphics[width=\textwidth]{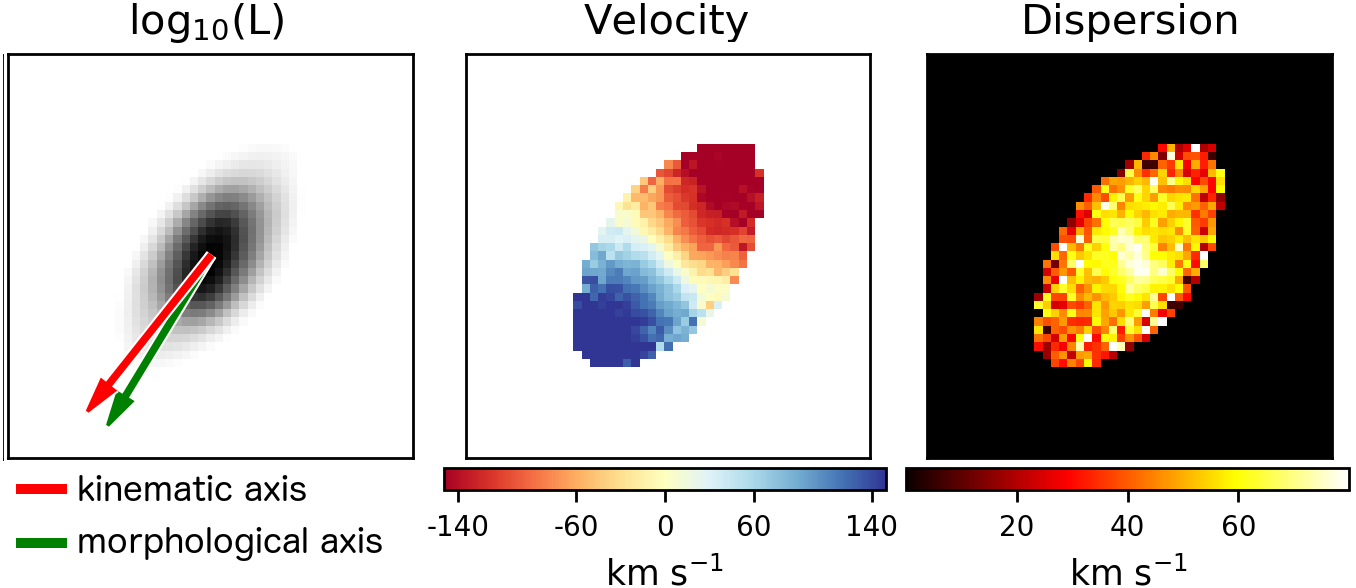}
    \caption{An example of our mock IFS observations for a kinematically aligned, oblate galaxy viewed at a random orientation. The maps are produced considering only stellar particles from the Illustris simulation. We have excluded spaxels with a low relative luminosity as these are often noisy and contribute little to integrated measurements such as $\lambda_{R_{e}}$. From left to right we show the stellar luminosity, stellar velocity, and stellar velocity dispersion maps. In the left panel we indicate the measured kinematic and morphological major axes, from which $\Psi$ is measured, with red and green arrows, respectively. The velocity and velocity dispersion maps produced by our kinematic mapping procedure are reminiscent of those found in IFS observations, in particular we reproduce the effect of beam smearing in the centre of the velocity dispersion map. The sharp edge in the stellar kinematics maps show the extent imposed by the imposed artificial detection limits.}\label{fig:kmap_ex}
\end{figure*}

We show an example of our output stellar luminosity and kinematic maps in Figure \ref{fig:kmap_ex}. In regards to this study, the key parameter measured from our kinematic maps is the kinematic position angle (see Section \ref{section:kinemis}). To accurately extract this quantity we simply require our maps to capture the rotation direction of each galaxy projection, thus in each artificial spaxel we must reliably recover the centroid of the LOSVD. A minor caveat here is that we have not incorporated the effects of spectral resolution in our mock LOSVD (e.g. convolving each LOSVD with a Gaussian comparable to the SAMI spectral resolution prior to fitting), but note that for each spaxel this will only affect $\sigma_*$ and not the centroid value. Differences in $\sigma_{*}$ for each projection will affect individual measurements of $\lambda_{R_e}$ and will only factor into our selection of fast vs slow rotators when comparing to observed samples in \ref{section:sim_v_obs}. Beyond the intrinsic stellar kinematics of a given galaxy, the classification as fast or slow rotator is more strongly dependent on the inclination of individual projections than on the details of our kinematics fitting. Thus, we expect no appreciable effect on our results due to the exclusion of spectral resolution in our analysis.

Finally, we note that in the case of stellar luminosity maps we have convolved with the seeing after producing an non-smoothed stellar luminosity map while the effects of seeing on our kinematics maps are incorporated during the construction of the maps. While not fully consistent, we have performed a test by comparing measured quantities from our luminosity maps, ellipticity ($\epsilon$) and kinematic misalignment ($\Psi$, see Section \ref{section:kinemis}), between maps produced as described above and maps produced in a similar way to our kinematics maps. This test involves producing maps using both methods at 500 different random projections, measuring $\epsilon$ and $\Psi$ for both maps, and recording the difference between the two methods. We find no systematic difference when comparing the two methods. Fitting a Gaussian to the difference distributions for $\epsilon$ and $\Psi$ we find $\sigma(\epsilon_{1}-\epsilon_{2})\simeq 0.02$ and $\sigma(\Psi_{1}-\Psi_{2}) \simeq 1^{\circ}$. Although there is some variation with galaxy type, we find similar values in a handful of galaxies tested. Finally we note that projections with the largest differences in $\Psi$ result from nearly circular projections where the photometric position angle is poorly defined, thus having a large error regardless of method. From these tests we conclude that, in practice, differences in smoothing implementation between stellar luminosity and kinematics maps will have no appreciable effect on our results.

\subsubsection{2D Kinematic Misalignment}\label{section:kinemis}

The projected kinematic misalignment angle ($\Psi$) is defined as 
\begin{equation}\label{eq:psi}
\sin{\Psi}=|\sin(PA_{\rm phot}-PA_{\rm kin})|,
\end{equation}
following \citet{Franx91} where $PA_{\rm phot}$ and $PA_{\rm kin}$ are the photometric and kinematic position angles, respectively. Thus for each 2D projected mock observation we must measure these two position angles, giving a mock value for the observed $\Psi$.

We measure $PA_{\rm phot}$ using the source finding code \textsc{ProFound} \citep[][https://github.com/asgr/profound]{Robotham18}. \textsc{ProFound} is a source finding algorithm executed on dilated segmentation maps encompassing the entire flux of the source galaxy (or star). Running \textsc{ProFound} on our $r$-band stellar luminosity maps provides measures of ellipticity ($\epsilon$) and effective (half-mass) radius $r_{e}$ in addition to $PA_{\rm phot}$. Although the measured value of $\epsilon$ is sensitive to the effects of seeing, here we do not attempt to correct for this. We also estimate the error in $\epsilon$ and $PA_{\rm phot}$ by evaluating these two values in a range radii encompasing 40-60\% of the total luminosity and taking the minimum and maximum as the lower and upper error. We note, however, that in most cases the errors on $\Psi$ are smaller than the smoothing kernel used in our fitting procedure described in Section \ref{section:shapefitting}, which results in overall poor fits to the data. Thus, in practice we simply assume a constant error of 5$^{\circ}$, which matches the procedure used in shape recovery of SAMI data \citep{Foster17}. 

We note here that in \citet{Foster17}, $\epsilon$ of SAMI galaxies are not measured from the SAMI datacubes, but from ancilliary photometric data with better seeing than the $\sim$2$\farcs$06 of SAMI (typically from SDSS, median seeing of 1$\farcs$43) while here we measure $\epsilon$ and kinematic values assuming the same seeing. We have tested the effect of higher spatial resolution on our $\epsilon$ measurement finding that, in general, $\epsilon$ decreases with decreasing spatial resolution. This effect is most pronounced at high $\epsilon$, however we do not expect this to have a large impact on our comparison to observations. Even at the highest $\epsilon$ we find a reduction in $\epsilon$ of only $\sim$0.05-0.07 when comparing $\epsilon$ measured assuming median SDSS vs median SAMI seeing for the same projection of a given galaxy. We also tested variation in $PA_{\rm phot}$ with spatial resolution finding typical variation of $<2^{\circ}$. For projections with very low $\epsilon$, $PA_{\rm phot}$ may vary more significantly, but this is due to the fact that $PA_{\rm phot}$ is poorly defined for nearly round projections, an effect also seen in observations. Given that we find such a small effect of decreased spatial resolution on $\epsilon$, we do not attempt to correct for this in our analysis.

$PA_{\rm kin}$ is measured using kinemetry on mock stellar velocity maps similarly to the method outlined in Appendix C of \citet{Krajnovic06}. Briefly, this is done by gradually rotating the mock velocity map and producing a bi-anti-symmetric velocity map, $V'(x,y)$  defined as:
\begin{equation}
	V'(x,y) = \frac{V(x,y)+V(x,-y)+V(-x,y)+V(-x,-y)}{4}
\end{equation}
where the origin is placed at the centre of the velocity map. Maps produced in Section \ref{section:mocks} are shifted based on the median of all particle positions such that the galaxy is centred at $x=y=z=0$, and the map size is specified such that there are an odd number of pixels on a side. This latter specification ensures that velocity maps have an unambiguous central pixel. At each rotation angle we calculate the mass weighted $\chi^{2}$ as:
\begin{equation}
	\chi^{2} = \sum_{i}^{N} \left( \frac{(V'_{i}-V_{i})(M_{*,i})}{M_{*,i}}\right)^{2}
\end{equation}
and identify $PA_{\rm kin}$ as the angle that minimises $\chi^{2}$. 

Each Illustris galaxy is rotated and viewed at 50 random projections where projection viewpoint is sampled with spherical uniformity. At each viewpoint we measure $PA_{\rm phot}$ and $PA_{\rm kin}$ and compute $\Psi$ using Eq. \ref{eq:psi}. In addition, the values of $\epsilon$ and $r_{e}$ allow us to also compute $\lambda_{R_e}$, which is commonly used as a proxy for the galaxy spin parameter, as \citep{Emsellem11}:
\begin{equation}
 \lambda_{R_e} = \frac{\sum_{i}^{N} M_{*,i}R_{i}|V_{i}|}{\sum_{i}^{N} M_{*,i}R_{i}\sqrt{V_{i}^{2}+\sigma_{i}^{2}}}
\end{equation}
where $R_{i}$ is the ellipsoidal radius of the $i$th pixel in our mock stellar luminosity and stellar kinematics maps. We are thus able to explore the position of each projection in both $\Psi$-$\epsilon$ and $\epsilon$-$\lambda_{R_e}$ space and compare directly with observations \citep[e.g.][]{Foster17}.

\subsection{Recovering the intrinsic shape}\label{section:shapefitting}
We use the algorithm of \citet{Foster17} to invert the distributions of the mock observables: the apparent $\epsilon$ and $\Psi$. The algorithm applies the inversion described by \citet{Franx91} to carefully selected sub-samples.

The observables ($\epsilon$ and $\Psi$) may be rewritten as a function of the intrinsic shape parameters ($p$ and $q$), the $\Psi_{\rm int}$, and the line-of-sight projection angles in spherical coordinates $0\le\varphi\le\pi$ and $0\le\nu\le2\pi$ \citep{Contopoulos56}. Hence, we obtain the following mathematical dependencies:
\begin{equation}\label{eq:PsiEps}
\Psi=\Psi(\Psi_{\rm int}, p, q, \varphi, \nu);\ {\rm and} \
\epsilon=\epsilon(p,q,  \varphi, \nu).
\end{equation}

Following \citet{Contopoulos56}, we rewrite the observed ellipticity (and eccentricity, $e$) as a function of the intrinsic axis ratios and projection angles:
\begin{equation}
e=(1-\epsilon)^2=\frac{a-\sqrt{b}}{a+\sqrt{b}},
\end{equation}
where
\begin{multline}
a=(1-q^2)\cos^2{\nu}+(1-p^2)\sin^2{\nu}\sin^2{\varphi}+p^2+q^2,\\
b=[(1-q^2)\cos^2{\nu}-(1-p^2)\sin^2{\nu}\sin^2{\varphi}-p^2+q^2]^2+\\
4(1-p^2)(1-q^2)\sin^2{\nu}\cos^2{\nu}\sin^2{\varphi}.
\end{multline}

We further define the triaxiality parameter $T$ as per \citet{Franx91}:
\begin{equation}\label{eq:triax}
T=\frac{1-p^2}{1-q^2}.
\end{equation}
Oblate ($a=b$) and prolate ($b=c$) systems have $T=0$ and $T=1$, respectively. Intermediate values of $T$ indicate triaxial systems in which no two principal axes lengths agree ($a\ne b\ne c$). The projected kinematic position angle depends on the line-of-sight angles and the intrinsic kinematic misalignment as follows:
\begin{equation}\label{eq:pakin}
\tan{(PA_{\rm kin})}=\frac{\sin\varphi\tan\Psi_{\rm int}}{\sin\nu-\cos\varphi\cos\nu\tan\Psi_{\rm int}}.
\end{equation}
The projected photometric position angle can be re-written as a function of $(T, \varphi, \nu)$ via the projection matrix as follows \citep{deZeeuw89}:
\begin{equation}\label{eq:pamin}
	\tan(2PA_{\rm min}) =\frac{2T\sin\varphi\cos\varphi\cos\nu}{\sin^2\nu-			T(\cos^2\varphi-\sin^2\varphi\cos^2\nu)},
\end{equation}
where $PA_{\rm min}=PA_{\rm phot}+\pi/2$ is the position angle of the projected short axis. Equations \ref{eq:pakin} and \ref{eq:pamin} are then combined with Equation \ref{eq:psi}.

We assume that $p$ is log-normally distributed $Y=\ln{(1-p)}$ with mean $\mu_Y$ and standard deviation $\sigma_Y$ \citep{Padilla08,Weijmans14,Foster17}. For $q$, we assume a normal distribution with mean $\mu_q$ and standard deviation $\sigma_q$. By construction, every mock observation is a random projection of the three-dimensional ellipsoid. This allows us to marginalise over all viewing angles $(\varphi,\nu)$. The probability of each set of viewing angles is the ratio of the area element over the total area of the unit sphere of viewing angles:

\begin{equation}
 P(\varphi,\nu)=\frac{\sin\nu}{4\pi}.
\end{equation}

In order to eliminate one further unknown parameter, we assume that $\Psi_{\rm int}$ depends on the intrinsic shape alone. We initially follow \citet{Weijmans14} and assume that $\Psi_{\rm int}$ coincides with the viewing direction that generates a round apparent ellipticity:
\begin{equation}\label{eq:thetaint}
 \tan(\Psi_{\rm int})=\sqrt{\frac{T}{1-T}}.
\end{equation}
Under this assumption, $\Psi_{\rm int}$ is 0$^{\circ}$ and 90$^{\circ}$ for perfectly oblate and prolate systems, respectively (see Fig. \ref{fig:Psi_T}). Values of $\Psi_{\rm int}$ quickly depart from 0$^{\circ}$ and 90$^{\circ}$ for triaxial systems \citep[see][their appendix A, for justification]{Weijmans14}.

We fit the intrinsic shape parameters ($\mu_Y,\sigma_Y,\mu_q,\sigma_q$), following the method of \citet{Foster17}. We minimise the square of the area between the modeled $F_{\rm mod}$ and the observed $F_{\rm obs}$ normalized distributions for $\Psi$ and $\epsilon$:

\begin{dmath}\label{eq:chisq} 
A^2 =  \sum_{i} (F_{\rm obs}(\Psi_{i})-F_{\rm mod}(\Psi_{i}))^2 (\delta \Psi_{i})^{2} + \\ \sum_{j} (F_{\rm obs}(\epsilon_{j})-F_{\rm mod}(\epsilon_{j}))^2 (\delta \epsilon_{j})^{2}.
\end{dmath}

We use the R package {\sc DEoptim} to efficiently minimize Equation \ref{eq:chisq} using differential evolution \citep[see][for more detail]{Mullen11}. We set a maximum of 200 iterations for each sub-sample with a threshold at $A^2 < 0.0003$ (i.e. total area of $A \lesssim 0.02$). We choose generous variable bounds: $-7\le\mu_Y\le0$, $0\le\sigma_Y\le7$ and $0\le\mu_q\le1$, $0\le\sigma_q\le1$.

\section{Results}\label{section:results}

\begin{figure*}
	\includegraphics[width=\textwidth]{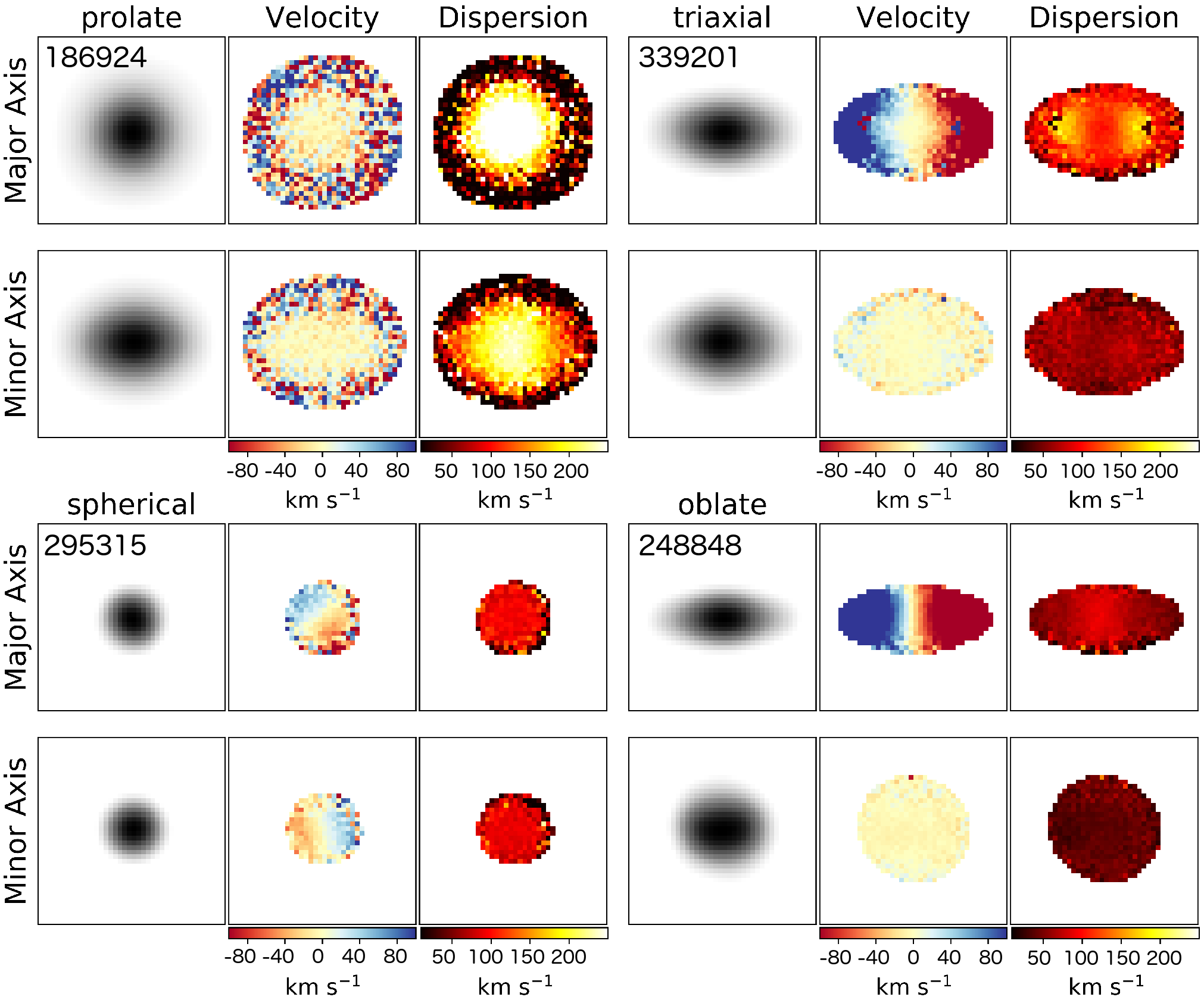}
    \caption{Representative examples of each shape class. For each galaxy we show from left to right the stellar luminosity, stellar velocity, and stellar velocity dispersion maps. The Illustris ID of the halo containing each galaxy is indicated in the top left of each luminosity map.}\label{fig:ex_maps}
\end{figure*}

\subsection{What shapes galaxies?}\label{section:what_shapes_galaxies}

\begin{figure*}
	\includegraphics[width=\textwidth]{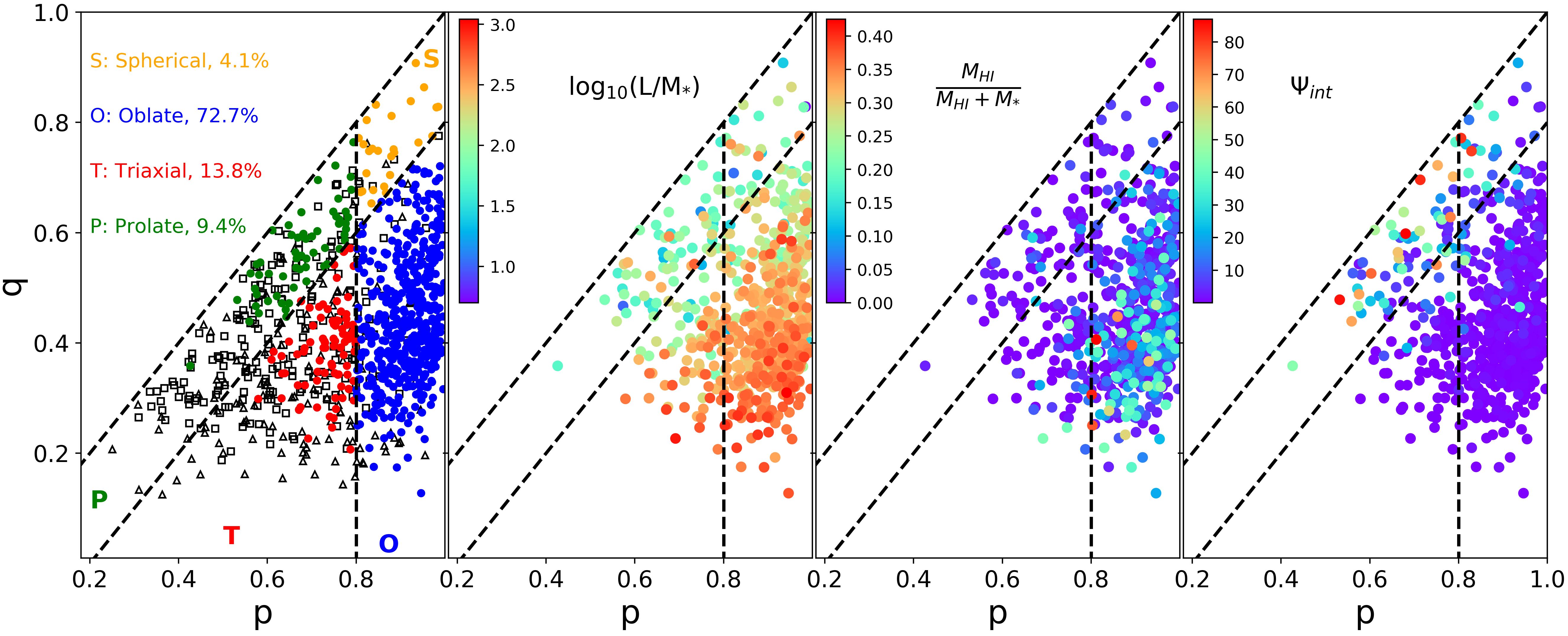}
    \caption{p vs q distribution for our final sample. The left panel illustrates the definition of spherical, oblate, triaxial, and prolate taken from \citet{Li18b}. We also show in the left panel barred and disturbed galaxies with open black squares and triangles, respectively, which do not pass our morphology cuts. The remaining panels show the same distribution, with the exception of barred and disturbed galaxies, with points colour coded based on other galaxy properties. From left to right, starting with the second panel, these properties are specific angular momentum ($L/M_{*}$), atomic gas fraction ($M_{HI}/M_{HI}+M_{*}$), and $\Psi_{\rm int}$, respectively.}\label{fig:pq_cbars}
\end{figure*}
 
First we explore the relationship between galaxy shape and other intrinsic galaxy properties in order to understand what key factors determine the shape of a given galaxy. Following \citet{Li18b} we separate galaxies into four basic shapes based on the axis ratios of their best fitting ellipsoids. This is shown in the top left panel of Figure \ref{fig:pq_cbars}. Galaxies are classified as spherical, oblate, prolate, and triaxial as follows:
\begin{multline}
 \textrm{Spherical:} \quad p-q < 0.2 \quad \& \quad p \geq 0.8\\
 \textrm{Oblate:} \qquad p-q \geq 0.2 \quad \& \quad p \geq 0.8\\
 \textrm{Prolate:} \quad \; \ \ p-q < 0.2 \quad \& \quad p < 0.8\\
 \textrm{Triaxial:} \quad \; \  p-q \geq 0.2 \quad \& \quad p < 0.8\\
\end{multline}
Example stellar luminosity and stellar kinematics maps for galaxies representative of each class are shown in Figure \ref{fig:ex_maps}. We find that the majority (72.7\%) of massive ($M_{*} \gtrsim 10^{10}$) Illustris galaxies are oblate systems.

We have also highlighted in the left panel barred and disturbed galaxies using open black squares and triangles respectively. Here we have identified barred and disturbed galaxies by visually inspecting three stellar luminosity maps of each halo (one projection for each ellipsoidal axis) produced as outlined in Section \ref{section:mocks}. Here, bars are identified as thin structures seen in face-on projections that are clearly embedded within a more extended disk. We find that many barred galaxies masquerade as prolate galaxies indicating that our 3D shape recovery technique based on the reduced inertia tensor is returning the shape of the bar rather than the galaxy as a whole. Disturbed galaxies are identified by large offsets between the luminosity peak and the luminosity weighted galaxy centre (central pixel of our luminosity maps), halos with multiple strong luminosity peaks, galaxies with prominent tails, or a combination of these. Disturbed galaxies often have low measured $q$ and a wide range of $p$. This is because long extensions (i.e. tidal tails) or widely separated subcomponents in a single galaxy (i.e. infalling galaxies) are interpreted as overly flat distributions by our fitting algorithm. We remove barred and distrubed galaxies from the remaining panels for clarity.

While methods to separate galaxies from cosmological simulations identified with the same halo, such galaxies often exhibit asymmetries that can be problematic for accurate ellipsoidal fitting. Thus, we simply exclude disturbed galaxies from the remaining panels of Figure \ref{fig:pq_cbars} for clarity. We note that barred and disturbed galaxies are included in our shape recovery tests presented in Section \ref{section:altass}, as such galaxies can not always be reliably removed from observed samples, depending on orientation. In Section \ref{section:altass} we also check the effects on our results of omitting barred and disturbed galaxies finding no appreciable differences.

In the remaining panels of Figure \ref{fig:pq_cbars} we also plot the $p$-$q$ distribution for massive galaxies in our sample, but colour points based on various galaxy properties. In the second panel from the left, we colour points by specific angular momentum ($L/M_{*}$) in units of kpc km s$^{-1}$. We find that $L/M_{*}$ exhibits the most striking correlation with galaxy shape. We compute a Pearson correlation coefficient between $L/M_{*}$ and $q$ of -0.60 with a p-value of $1.7\times10^{-63}$. This means that, at fixed mass, more rapidly rotating galaxies are more flat. The third panel shows the atomic gas fraction, $M_{HI}/(M_{HI}+M_{*})$, clearly demonstrating that the majority of gas rich galaxies are oblate in shape. Finally, we show $\Psi_{\rm int}$ in the last panel, indicating that the majority of galaxies with kinematic offsets are either prolate or spherical.
In the remainder of this section we examine the ensemble properties of each subset in galaxy shape in order to better understand the processes that result in each galaxy type.
\vspace{.1in}

\noindent \textbf{Spherical Galaxies:}

Spherical galaxies make up the smallest subset (4.1\%) of galaxies represented in Figure \ref{fig:pq_cbars}. Spherical galaxies are typically massive and gas poor with a median log$_{10}(M_{*}) = 11.2$ M$_{\odot}$ and a median neutral hydrogen mass fraction of 0.01. The highest neutral gas fraction for an individual spherical galaxy is 0.17, though this galaxy sits on the line separating spherical galaxies from oblate. We find a wide range of $\Psi_{\rm int}$ values for spherical galaxies, ranging from 0.9$^{\circ}$ to 81.2$^{\circ}$. We note, however that for truly spherical galaxies ($p=q=1$), $\Psi_{\rm int}$ can not be reasonably quantified as the major axis direction is undefined. In practice, for galaxies in hydrodynamical simulations, this means that galaxies near the upper-right corner of Figure \ref{fig:pq_cbars} the major axis direction is dependent on the stochastic positions of particles in any given timestep. For this reason, we do not put much emphasis on the value of $\Psi_{\rm int}$ for spherical galaxies when considering galaxies with significant kinematic offsets.

\vspace{.1in}
\noindent \textbf{Oblate Galaxies:}

The majority (72.7\%) of galaxies in the $z=0$ snapshot of the Illustris-1 simulation are found to be oblate. Typically, oblate galaxies are slightly lower in mass (median log$_{10}(M_{*}) = 10.97$ M$_{\odot}$) compared to spherical galaxies, though there is significant variation. 

One defining feature of many (but not all) oblate galaxies is the presence of neutral gas. We  find that neutral gas fraction in oblate galaxies varies from 0.0 to 0.43, with more gas rich galaxies typically found at lower $q$. Considering relatively gas rich galaxies with $M_{H}/(M_{H}+M_{*}) > 0.05$ of all shapes, we find that 87.4\% (159/182) are classified as oblate with majority of the remainder (16/23) being triaxial. This is not surprising as galaxies with large gas fractions at $z=0$ will have recently accreted this gas. Recently \citet{Garrison-Kimmel18} showed that gas accretion at low redshift in hydrodynamical zoom simulations has a higher impact parameter than accretion occurring at high redshift, often resulting in a disk distribution. A similar conclusion is reached by \citet{Bryant18} who show that samples of late-type galaxies from the SAMI survey show a conspicuous lack of examples with gas rotating out of the disk plane. 
Thus, stars that form from recently accreted gas in late-type galaxies will inherit this disk distributions explaining why the majority of gas-rich galaxies at $z=0$ in Illustris are oblate.

\vspace{.1in}
\noindent \textbf{Triaxial Galaxies:}

Triaxial galaxies make up an appreciable fraction of our sample at 13.8\%. Triaxial galaxies share similar properties to oblate galaxies being slightly more massive, having a median log$_{10}(M_*) = 11.1$ M$_{\odot}$, though, again, with significant variation. We also find a similar range in neutral gas fraction from 0.0 to 0.22, though a larger fraction of triaxial galaxies are gas poor when compared to oblate galaxies. Overall, triaxial galaxies appear in Figure \ref{fig:pq_cbars} as an extension of the oblate population to lower $p$. The lower typical gas fraction and larger $M_*$ of triaxial galaxies when compared to oblate galaxies suggests that the consumption of neutral gas in oblate galaxies may be related to structural changes.

\vspace{.1in}
\noindent \textbf{Prolate Galaxies:}

Similar to spherical galaxies, prolate galaxies are relatively massive and gas poor. We find a median stellar mass of log$_{10}(M_{*}) = 11.3$ M$_{\odot}$, and a gas fraction in the range 0.0-0.13 (median = 0.006). Prolate galaxies have a slightly larger representation than spherical galaxies however, making up 9.4\% of the total sample. 

The most interesting characteristic of prolate galaxies is the fact that the majority of galaxies with significant kinematic offsets are found to be prolate. Considering all galaxies with $\Psi_{\rm int} > 30^{\circ}$ we find that 67\% (24/36) are prolate, 19\% (7/36) are spherical, 8\% (3/36) are triaxial, and 6\% are oblate (2/36). We have already mentioned that defining morphological axes for spherical axes is often degenerate, thus $\Psi_{\rm int}$ for this subsample should be considered with measured skepticism. Considering kinematically offset triaxial and oblate galaxies we find these are typically found towards the high-$q$ end of the selection box, i.e. more spherical/prolate. \citet{Li18b} have shown that prolate galaxy shapes in Illustris are typically associated with major, dry mergers. If kinematically offset prolate galaxies are also formed in this manner, this may suggest that offset kinematics are a direct result of mergers and, likely, dependent on merger geometry. We discuss the role of mergers in observed kinematic offsets further in Section \ref{section:mergerhist}. Finally, we point out that the idea that $\Psi_{\rm int} = 90^{\circ}$ is a defining characteristic of prolate galaxies \citep[i.e.][]{Tsatsi17} is untrue. 

\vspace{.3in}

Before moving on, we summarise here the salient points of our initial assessment of the properties of galaxies of different shape classes. 

The lower mass regime of our sample (log$_{10}(M_*) \simeq 10-11$ M$_{\odot}$) is dominated by oblate and triaxial systems. We also find that nearly all gas rich ($M_{H}/(M_{H}+M_{*}) > 0.05$) systems are one of these two types, with 87.4\% being classified as oblate. These results point to a scenario in which neutral gas accreted at low redshift typically settles into a thin, rotating disk, thus leaving behind a flattened disk of stars. The relative number of gas rich oblate and triaxial galaxies may further suggest that gas consumption in oblate galaxies is related to structural changes resulting in a reduction of $p$ and/or an increase in $q$. Clearly this does not happen in all cases, however, given the large number of gas poor oblate systems. 

Prolate and spherical galaxies are typically more massive and less gas rich. It is important to note that defining morphological axes for a perfectly spherical galaxy is a completely degenerate task. Thus, reliably defining $\Psi_{\rm int}$ for this class of galaxy is often not possible. Prolate galaxies, on the other hand, are perhaps the most interesting class in regards to this work. Our first major result is that prolate galaxies have a 67\% representation among galaxies with $\Psi_{\rm int} > 30^{\circ}$. In the context of \citet{Li18b}, who show prolate galaxies result from dry, major mergers, this suggests kinematic offsets are directly related to past mergers. The secondary result from our initial analysis is that prolate morphology is not synonymous with $\Psi_{\rm int} = 90^{\circ}$, which has been termed ``prolate rotation'' \citep{Emsellem14,Tsatsi17,Ebrova17}.

\subsection{Challenging common assumptions about the dependence of $\Psi_{ \rm int}$ on intrinsic shape}\label{section:Psi_T}

\begin{figure}
	\includegraphics[width=\columnwidth]{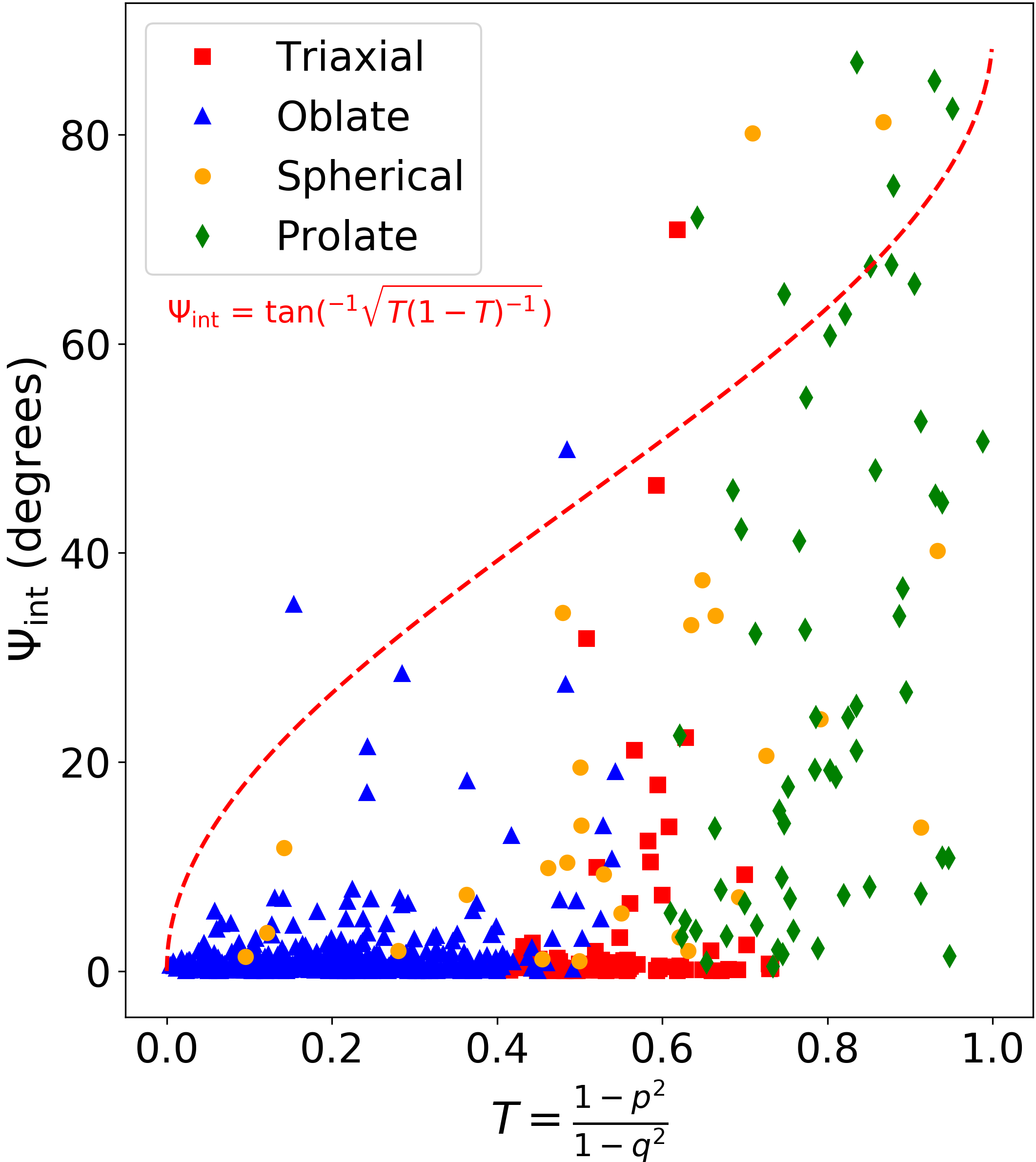}
    \caption{The distribution of galaxy triaxiliaty ($T$) vs $\Psi_{\rm int}$ for galaxies meeting our morphology cut (disturbed and strongly barred galaxies removed). Plotted symbols indicate shape subclasses defined in Section \ref{section:what_shapes_galaxies}. The red dashed line represents the relationship derived by \citet{Weijmans14} based on the theoretical work of \citet{Franx91} for elliptical galaxies with mass distributions described by St\"{a}ckel potentials. This relationship underpins some recent works attempting to recover galaxy shapes based on IFS observations \citep[e.g.][]{Foster17}. While the dashed line may represent a rough limiting case, particularly at low $T$, it is a poor description of galaxies from the Illustris simulation.}\label{fig:Psi_T}
\end{figure}

In this Section we test whether or not Illustris galaxies follow the relationship between shape and $\Psi_{\rm int}$ put forward by \citet{Weijmans14} based on the theoretical work of e.g. \citet{Franx91}, \citet{Hunter92}, and \citet{Arnold94}, which underpins some observational efforts aimed at recovering the shapes of galaxy samples from IFS observations \citep[e.g.][]{Foster17}. We show in Figure \ref{fig:Psi_T} the distribution in $T$ vs $\Psi_{\rm int}$ for galaxies in Illustris meeting our stellar mass and morphological cuts, ensuring that plotted galaxies have well defined values of $p$, $q$, and $\Psi_{\rm int}$. The relationship of \citet{Weijmans14} is also shown with a red dashed line. From Figure \ref{fig:Psi_T} it is clear that the assumption that $\Psi_{\rm int}$ is dependent on the triaxiality parameter does not seem to hold for Illustris galaxies. 

Figure \ref{fig:Psi_T} also shows that the vast majority of oblate and triaxial Illustris galaxies have $\Psi_{\rm int}\simeq0.0$. This is likely due to the fact that gas accretion at low redshift often occurs with a large impact parameter resulting in a rapidly rotating distribution that flattens into a thin disk \citep{Bryant18,Garrison-Kimmel18} with rotation about the minor axis (i.e. $\Psi_{\rm int} = 0.0^{\circ}$). By definition, galaxies classified as oblate or triaxial in this work are necessarily flattened having $p-q > 0.2$. Thus, our results suggest that gas accretion resulting in a kinematically aligned disk is the preferred formation scenario for flattened (i.e. low $q$) galaxies. We also note that oblate and triaxial galaxies having relatively large values of $\Psi_{\rm int}$ almost exclusively occupy the high $q$ regions of the oblate and triaxial class definition regions of Figure \ref{fig:pq_cbars}, consistent with this scenario.

Prolate and spherical galaxies, on the other hand are found to exhibit a wide range in $\Psi_{\rm int}$. In the case of spherical galaxies, this is partially driven by the stochasticity of the direction of the major morphological axis. In other words, the variation in the length of the axes over time due to the motions of particles may be larger than the difference between the individual axes. Prolate galaxies have a relatively well defined major axis, thus $\Psi_{\rm int}$ measurements are more reliable when compared to spherical galaxies. Considering the distribution of $\Psi_{\rm int}$ for prolate galaxies we find a slight excess of $\Psi_{\rm int} < 15^{\circ}$ galaxies, with the remaining galaxies being uniformly distributed. Thus, even for prolate galaxies with significant kinematic offsets there is no underlying relationship between shape and $\Psi_{\rm int}$ suggesting that kinematic offsets in galaxies result from random processes (see Section \ref{section:mergerhist}). 

The  relationship between $T$ and $\Psi_{\rm int}$ presented in Figure \ref{fig:Psi_T} is clearly overly simplistic. One reason for this is that, although we see a sequence in increasing $T$ from oblate to triaxial to prolate, spherical galaxies have a wide variation in $T$. This is the combination of two issues: first both $p$ and $q$ are close to 1 meaning that the numerator and denominator in Equation \ref{eq:triax} are small and second they are nearly equal to each other. The result being that small variations in $p$ vs $q$ result in large variations in $T$. Thus, galaxies with very different intrinsic shapes occupy the same regions of Figure \ref{fig:Psi_T}. Indeed, idealised spherical ($p=q=1$) and oblate disk ($q \ll p = 1$) would have the exact same value of $T=0$. The theoretical work underpinning the $T$-$\Psi_{\rm int}$ relationship given by the red dashed line in Figure \ref{fig:Psi_T} also predicts that, for idealised prolate galaxies in St\"{a}ckel potentials, the orbital family of ``box orbits" should be completely absent \citep{deZeeuw85}. Recently, however, \citep{wang19} showed by integrating particle orbits in Illustris galaxies that, at small radii, prolate-triaxial galaxies are actually dominated by box orbits. This highlights the fact that simple, static potentials may not always be applicable to real galaxies with ever changing gravitational potentials.

As we have seen, Illustris galaxies do not exhibit idealised shapes, which highlights the core issue with the $T$-$\Psi_{\rm int}$ relationship in Figure \ref{fig:Psi_T}: any departure from $T=0$ (i.e. $p \neq 1$) is expected to result in a rapid departure from $\Psi_{\rm int}=0$. Clearly this is not the case. Any improved assumption that connects kinematic offsets with intrinsic galaxy shape must allow for $\Psi_{\rm int}=0$ for a larger variety of combinations of $p$ and $q$ (see Section \ref{section:altass}).

\subsection{The Dependence of $\Psi_{\rm int}$ on Merger History}\label{section:mergerhist}

\begin{figure*}
	\includegraphics[width=17cm]{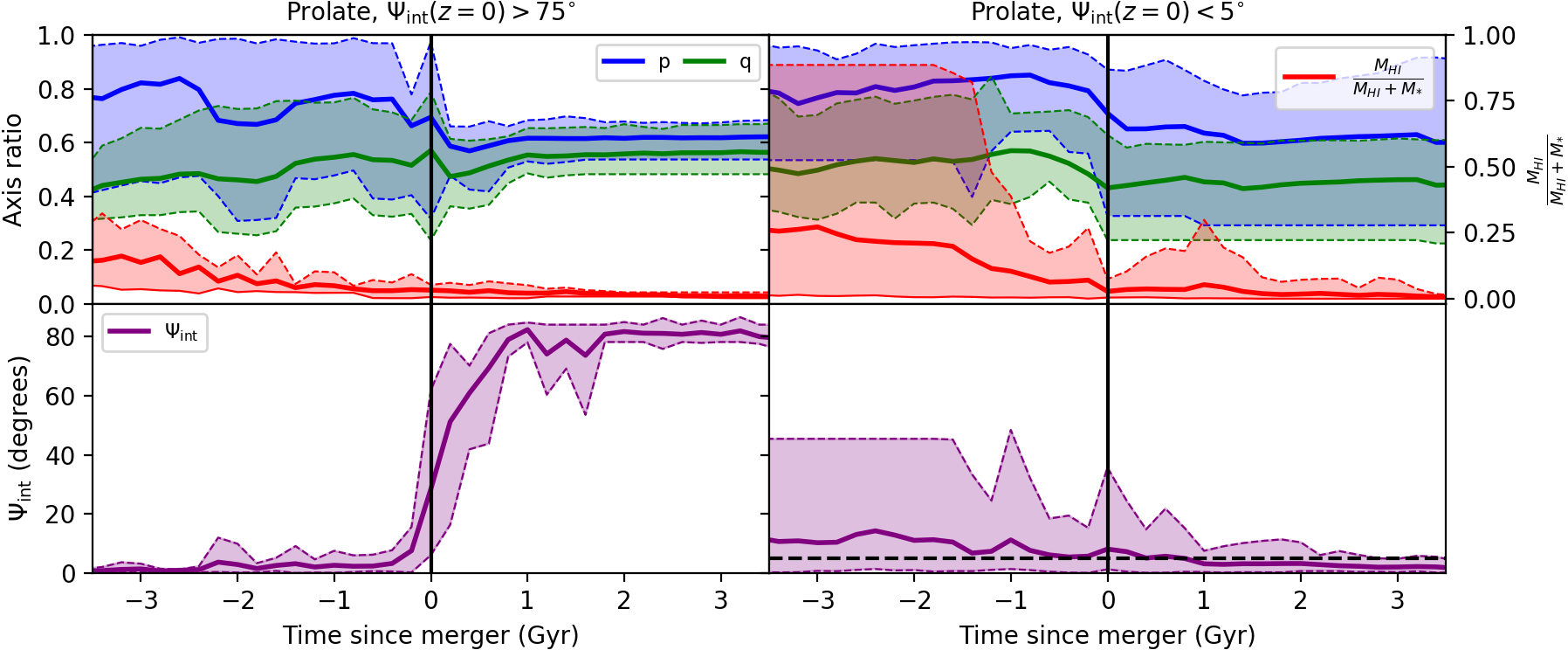}
    \caption{The time evolution of shape and $\Psi_{\rm int}$ for prolate galaxies where galaxies with $\Psi_{\rm int}>75^{\circ}$ and $\Psi_{\rm int} < 5^{\circ}$ at $z=0$ are shown in the left and right panels, respectively. Solid lines show the mean value of each subsample and dashed lines show in the minimum and maximum values and all have been smoothed with a Gaussian kernel having $\sigma=0.1$ Gyr. The time axis of each galaxy has been shifted such that 0 represents the time of its most recent major merger with negative values indicating times pre-merger. Top panels show evolution of $p$ (blue), $q$ (green), and neutral gas fraction (red) while bottom panels show the evolution of $\Psi_{\rm int}$. In both cases mergers are related to a change in shape caused by a reduction in $p$, however where $\Psi_{\rm int}(z=0)<2^{\circ}$ this change is more gradual on average. In the case of $\Psi_{\rm int}(z=0)>75^{\circ}$, this shape change is also directly associated with a flip in $\Psi_{\rm int}$ from $\sim0^{\circ}$ to $\sim90^{\circ}$. The other clear difference between these morpho-kinematic subsamples is $\Psi_{\rm int}(z=0)<2^{\circ}$ galaxies are often hosting significantly more $HI$ gas before and after the merger compared to $\Psi_{\rm int}(z=0)>75^{\circ}$ galaxies (though examples of kinematically aligned, prolate galaxies formed from dry mergers are also present).}\label{fig:mhist}
\end{figure*}

We have shown in the Sections \ref{section:what_shapes_galaxies} and \ref{section:Psi_T} that kinematically offset galaxies ($\Psi_{\rm int} > 30^{\circ}$) most often have prolate shapes. Furthermore, prolate galaxies with kinematic offsets are found to have a uniform distribution in $\Psi_{\rm int}$ with no clear dependence on $p$ or $q$ (see also Section \ref{section:altass}) suggesting kinematic offsets in prolate galaxies result from random processes. The obvious candidates for such processes are galaxy mergers. Indeed, prolate shapes in Illustris galaxies have been recently shown by \citet{Li18b} to result from dry, major mergers. In this section, we address whether or not this merger driven morphological change is also related to the emergence of kinematic offsets in prolate galaxies. In addition, we check whether or not the shapes of kinematically aligned prolate galaxies are also the result of major mergers.

To explore the merger histories of Illustris galaxies we trace back the ``first progenitor'' for each galaxy at each time step back to redshift $\sim$9.5. This traces the most massive progenitor back in time, thus the primary branch of the merger tree. At each time step we then measure $p$, $q$, $\Psi_{\rm int}$, and the stellar and neutral hydrogen gas masses to explore the co-evolution of each of these quantities. We then identify the last major merger by finding the last timestep at which the galaxy has experienced a 20\% growth compared to the previous timestep. This allows us to check if kinematic offsets in galaxy subsamples is related to merger activity, and which properties of the merger most strongly influence the kinematic offset of the merger remnant.

First we check the merger histories of prolate galaxies having $\Psi_{\rm int} > 75^{\circ}$, which are the prolate galaxies closest to exhibiting major axis rotation. After removing barred and disturbed galaxies from our sample, we are left with four prolate galaxies with such extreme kinematic offsets. We plot the resulting average shape and kinematic evolution of these four galaxies in the left panels Figure \ref{fig:mhist}. Here, the x-axis values come from the snapshot lookback times where we have shifted the values such that the time of the most recent major merger occurs at $t=0$ Gyr and negative values represent times pre-merger. In the top panel the solid blue, green, and red lines represent the average histories of $p$, $q$, and $M_{HI}/(M_{HI}+M_{*})$, respectively. We find that the time of the merger is directly correlated with a change in galaxy shape manifested as a reduction in $p$ such that $p\simeq q$. We find that this shape transition is also synchronised with a shift in $\Psi_{\rm int}$ from $\sim$0$^{\circ}$ to $\sim$90$^{\circ}$. This merger-driven morpho-kinematic transition is completed in $\sim$1 Gyr, which is often found to be near the lower limit for merger timescales in cosmological simulations \citep[e.g.][]{Jiang08,Young18}.

Our finding that kinematic transitions in prolate galaxies are merger-driven is similar to the results of \citet{Ebrova17}. We note that \citet{Ebrova17} allow galaxies with an order of magnitude fewer particles ($N>10^{4}$) than in this work resulting in roughly an order of magnitude larger initial sample ($N=$7697 vs $N=$630). \citet{Ebrova17} identify 59 galaxies in Illustris with large kinematic offsets (0.8\%, though using a slightly different metric) while we find 4-5 with $\Psi_{\rm int} > 75^{\circ}$ (0.6-0.8\%, depending on whether spherical galaxies are included or not). Thus, statistically the numbers of kinematically offset galaxies are consistent. We note that gas fractions quoted in \citet{Ebrova17} are significantly larger than those presented here. This is likely due to the fact that gas fractions in \citet{Ebrova17} are total gas fractions while those quoted here are $HI$ only. Indeed, massive, prolate galaxies in Illustris can be found to harbour very massive gas reservoirs, however the majority of this gas is extremely hot and thus not likely to collapse into a rotating disk.

Next we explore the merger histories of kinematically aligned, $\Psi_{\rm int} < 5^{\circ}$, prolate galaxies. Again removing barred and disturbed galaxies from our sample we find a subsample of 12 galaxies. ``Prolate rotation'' or "prolate-like rotation" has been used to describe galaxies that rotate about their major axis \citep[i.e. $\Psi_{\rm int} = 90^{\circ}$,][]{Emsellem14,Ebrova17,Tsatsi17}, thus understanding the origin of prolate galaxies without kinematic misalignment is of particular interest. Indeed, given that we find 12 prolate galaxies with $\Psi_{\rm int} < 5^{\circ}$ and only 4 with $\Psi_{\rm int} > 75^{\circ}$, it would appear that kinematic alignment is preferred over misalignment (noting however that 43/59 prolate galaxies have $\Psi_{\rm int}$ falling between these two strict limits). Thus, assuming $\Psi_{\rm int} = 90^{\circ}$ is an intrinsic property of prolate galaxies is a misconception. 

The average shape and kinematic histories of kinematically aligned, prolate galaxies are shown in the right panels of Figure \ref{fig:mhist}. We find that, similar to extremely kinematically offset, prolate galaxies, aligned galaxies have a slight shape transition associated with their last major merger. In this case, however, the change in shape is more gradual and the final value of $p$ is, on average, slightly higher with a larger difference between $p$ and $q$. We also find that the merger is not associated with a rapid change in $\Psi_{\rm int}$. One difference between kinematically aligned and misaligned, prolate galaxies in Figure \ref{fig:mhist} is the fact that, often, before the most recent major merger, aligned galaxies are more gas rich. As we have shown in Section \ref{section:what_shapes_galaxies}, gas richness is typically associated with an oblate shape and $\Psi_{\rm int}\simeq 0^{\circ}$. Thus, major mergers resulting in prolate galaxies that are gas rich appear to preferentially result $\Psi_{\rm int} \simeq 0.0^{\circ}$ remnants. This is clearly not the only pathway for forming $\Psi_{\rm int} \simeq 0.0^{\circ}$, prolate galaxies as at least one example of a gas free merger is present in the $\Psi_{\rm int} < 5.0^{\circ}$ sample in the right panels of Figure \ref{fig:mhist}.

What, then, is the formation mechanism of kinematically aligned, prolate galaxies from dry mergers? It is known that the gas and stellar distributions in  merger remnants are undoubtedly influenced by the details of the merger orbit and orientation \citep[e.g.][]{Bassett17}. As pointed out by \citet{Li18b}, the time resolution of snapshots output in the Illustris simulation is too coarse to track the complex interraction histories of individual mergers, thus we can not comment further here. For more detailed discussion of the origins of minor-axis rotation in prolate galaxies see \citet{Naab14}, \citet{Ebrova15}, \citet{Ebrova17}, and \citet{Li18b}.

\section{Recovering Galaxy Shape from Projected Maps}\label{section:altass}

\begin{table*}
  \begin{center}
    \caption{Intrinsic shape fit results for various samples (column 1) and/or assumption for the intrinsic kinematic misalignment dependency on intrinsic shape (column 2). The sample size is given in column 3. Mean and standard deviations of the lognormal distribution of intrinsic disc circularity ($Y=\ln(1-p)$) are given in columns 4 and 5, respectively. Columns 6 and 7 are the mean and standard deviation of the fitted distribution of intrinsic flattening ($q$). When fitted as a free parameter, the mean and standard deviations of the intrinsic kinematic misalignment ($/Psi_{\rm int}$) are given in columns 8 and 9, respectively. The quality of the fit can be assessed through the $A^2$ value (column 10) with lower $A^2$ values corresponding to better fits.}
    \label{tab:table1}
    \begin{tabular}{l|l|l|l|l|l|l|l|l|l|l} 
      \textbf{Sample} & \textbf{$\Psi_{\rm int}$} & $N_{\rm gal}$ &  \textbf{$\mu_Y$} & \textbf{$\sigma_Y$} & \textbf{$\mu_q$} & \textbf{$\sigma_q$} & \textbf{$\mu_{{\Psi}_{\rm int}}$} & \textbf{$\sigma_{{\Psi}_{\rm int}}$}  & \textbf{$A^2$}\\
      (1) & (2) & (3) & (4) & (5) & (6) & (7) & (8) & (9) & (10) \\
      \hline\hline
Spherical & $\tan^{-1}(\sqrt{T(1-T)^{-1}})$ &  33 & -0.43 & 3.85 & 0.86 & 0.22 & NA & NA & 0.0037 \\
Oblate & $\tan^{-1}(\sqrt{T(1-T)^{-1}})$ & 513 & -4.43 & 3.32 & 0.46 & 0.14 & NA & NA & 0.0005 \\
Prolate & $\tan^{-1}(\sqrt{T(1-T)^{-1}})$ & 184 & -0.02 & 2.04 & 0.48 & 0.09 & NA & NA & 0.0034 \\
Triaxial & $\tan^{-1}(\sqrt{T(1-T)^{-1}})$ & 247 & -0.92 & 2.79 & 0.38 & 0.06 & NA & NA & 0.0010 \\
\hline
Spherical & Free & 33 & -5.96 & 0.14 & 0.77 & 0.17 & 0.8 & 38.8 & 0.0046 \\
Oblate & Free & 513 & -0.82 & 1.71 & 0.46 & 0.17 & 1.1 & 3.5 & 0.0010 \\
Prolate & Free & 184 & -0.72 & 0.95 & 0.52 & 0.10 & 16.1 & 89.9 & 0.0011 \\
Triaxial & Free & 247 & -0.02 & 1.71 & 0.40 & 0.09 & 0.8 & 27.2 & 0.0011 \\
\hline
Spherical cleaned & Free & 28 & -5.62 & 0.58 & 0.78 & 0.17 & 0.4 & 40.6 & 0.0052 \\
Oblate cleaned & Free & 380 & -1.11 & 2.15 & 0.46 & 0.17 & 0.0 & 6.1 & 0.0007 \\
Prolate cleaned & Free & 86 & -0.29 & 0.81 & 0.58 & 0.12 & 82.9 & 81.0 & 0.0019 \\
Triaxial cleaned & Free & 135 & -0.83 & 1.71 & 0.41 & 0.10 & 1.9 & 26.2 & 0.0008 \\
\hline
Fast rotators & $\tan^{-1}(\sqrt{T(1-T)^{-1}})$ & 784 & -2.3 & 3.71 & 0.41 & 0.14 & NA & NA & 0.0003 \\
Slow rotators & $\tan^{-1}(\sqrt{T(1-T)^{-1}})$ & 193 & -0.10 & 2.07 & 0.50 & 0.03 & NA & NA & 0.0038 \\
       \hline
    \end{tabular}
  \end{center}
\end{table*}


It is clear from Section \ref{section:Psi_T} that the relationship between $\Psi_{\rm int}$ and galaxy shape proposed by \citet{Weijmans14} is a poor representation of galaxies in the Illustris simulation. In this section we explore first if there is an alternative assumption for $\Psi_{\rm int}$ that gives a more accurate description of Illustris galaxies. Next we test how the previous assumption affects our recovery of galaxy shape using quantities measured from projected 2D stellar luminosity and stellar kinematics maps, as well as if an alternative assumption can improve this recovery.

The first step in improving our shape recovery is to reassess the assumption connecting $\Psi_{\rm int}$ and 3D shape. Considering the results of Section \ref{section:Psi_T}, illustrated in Figure \ref{fig:Psi_T}, the key will be producing a model that allows $\Psi_{\rm int} = 0$ for a more generous range of $p$ and $q$ combinations. After considering various parameterisations, we find that $\Psi_{\rm int}$ does not correlate well with intrinsic shape ($p$, $q$ or various parametrisation of those). Instead, we explicitly fit for $\Psi_{\rm int}$, assuming a Gaussian distribution truncated between 0 and 90 degrees with mean $\mu_{\Psi_{\rm int}}$ and standard deviation $\sigma_{\Psi_{\rm int}}$. For spherical and prolate galaxies, $\Psi_{\rm int}$ is essentially random, thus $\sigma_{\Psi_{\rm int}}$ can be arbitrarily large to accommodate a broad distribution.

\begin{figure*}
\begin{center}
\includegraphics[width=84mm]{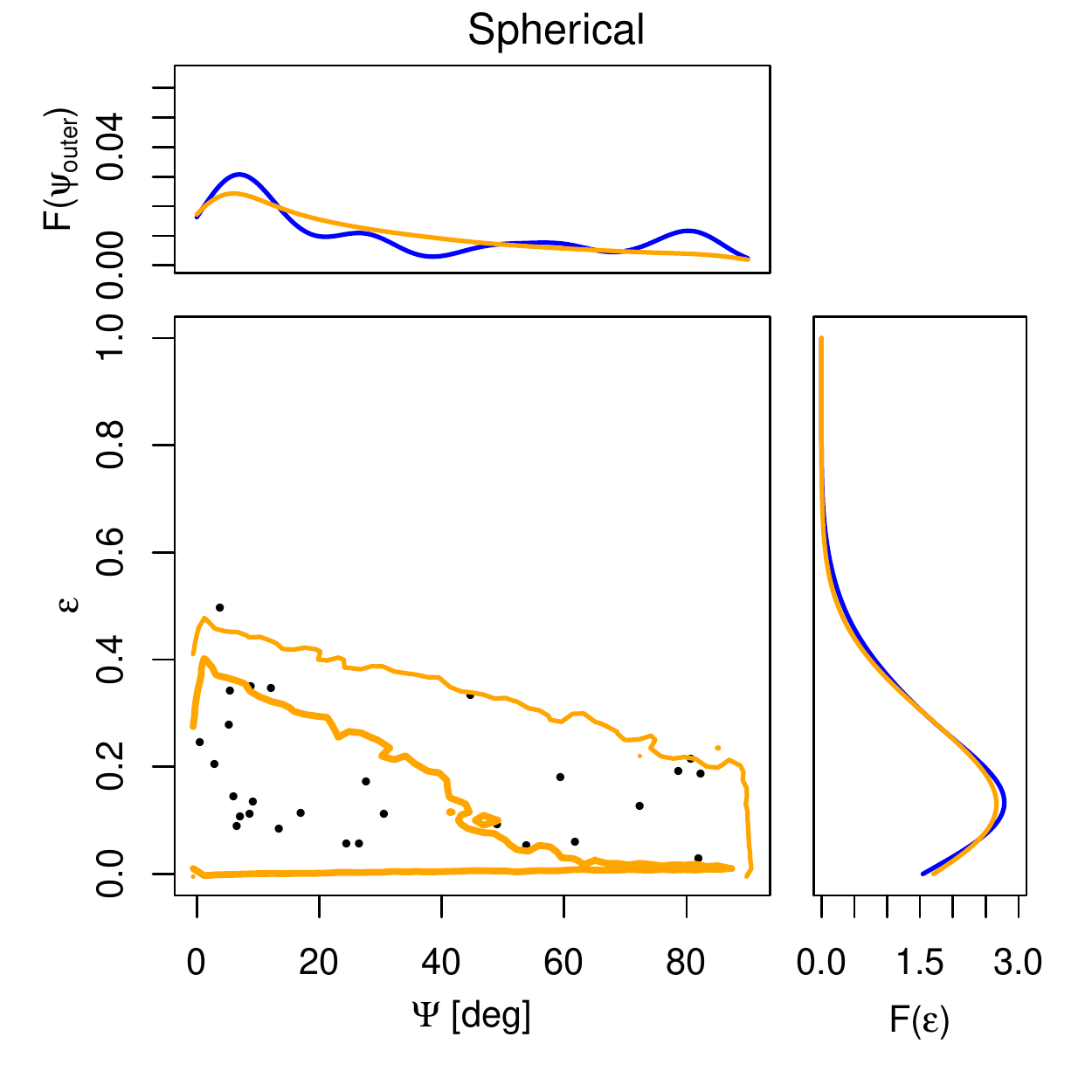} 
\includegraphics[width=84mm]{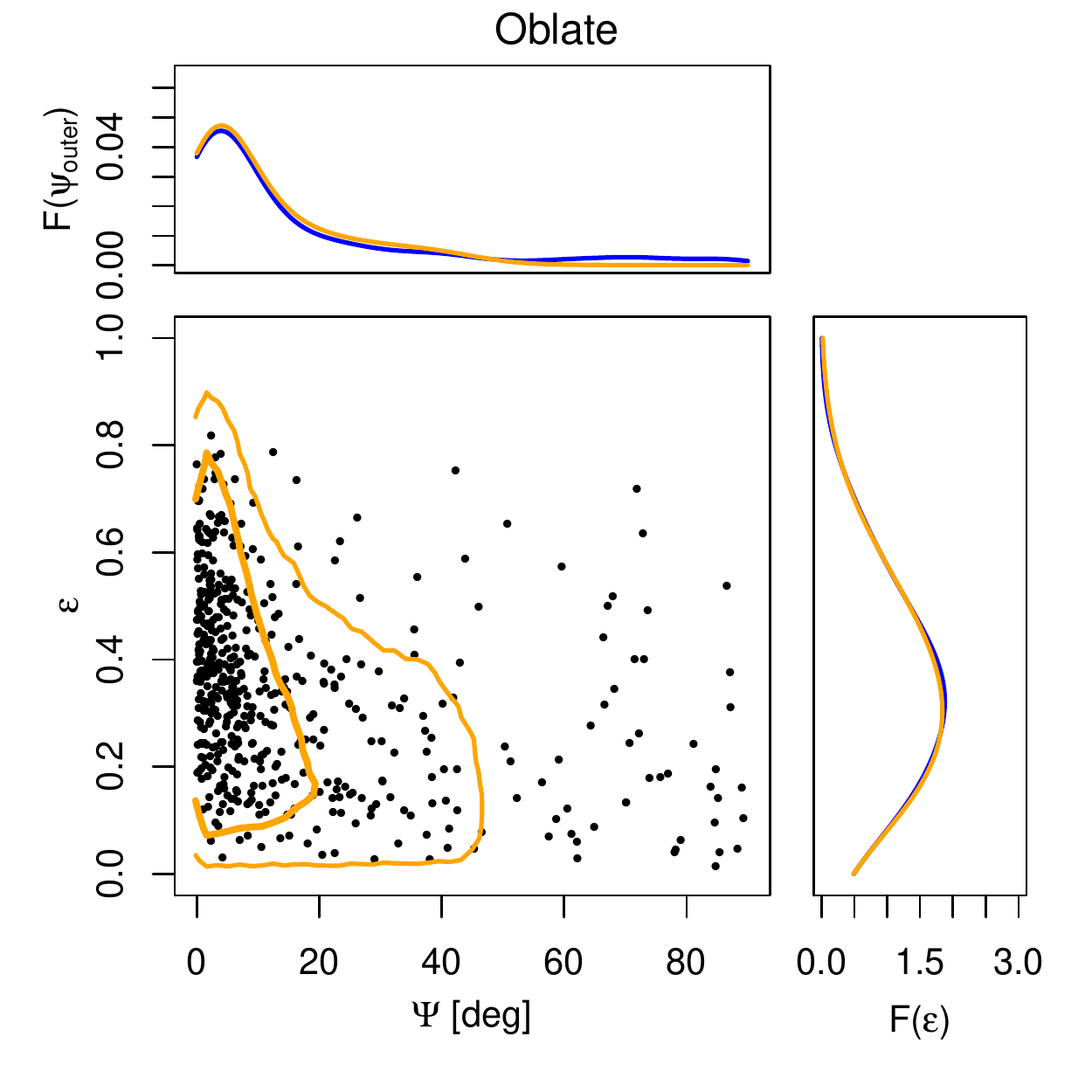}\\
\includegraphics[width=84mm]{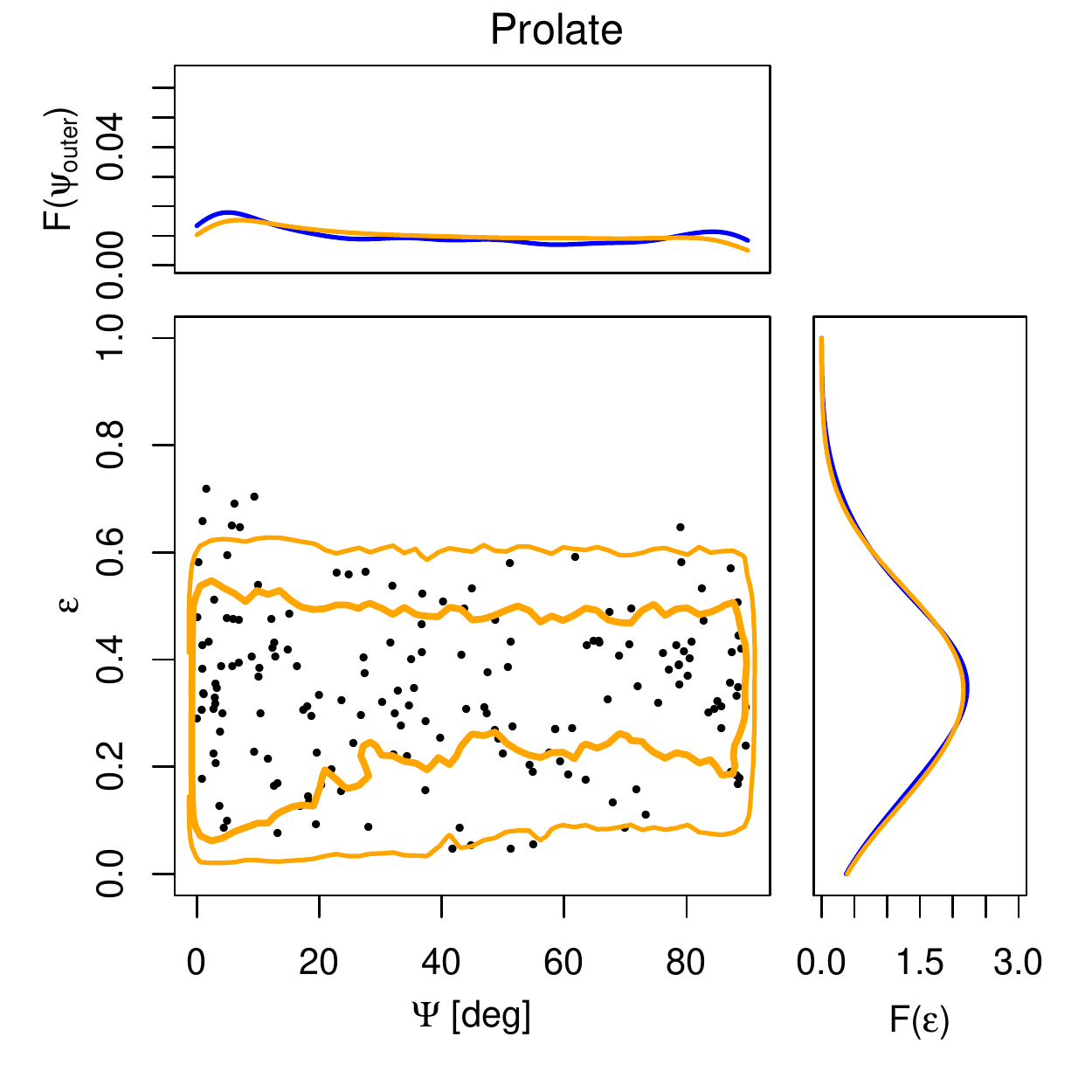}
\includegraphics[width=84mm]{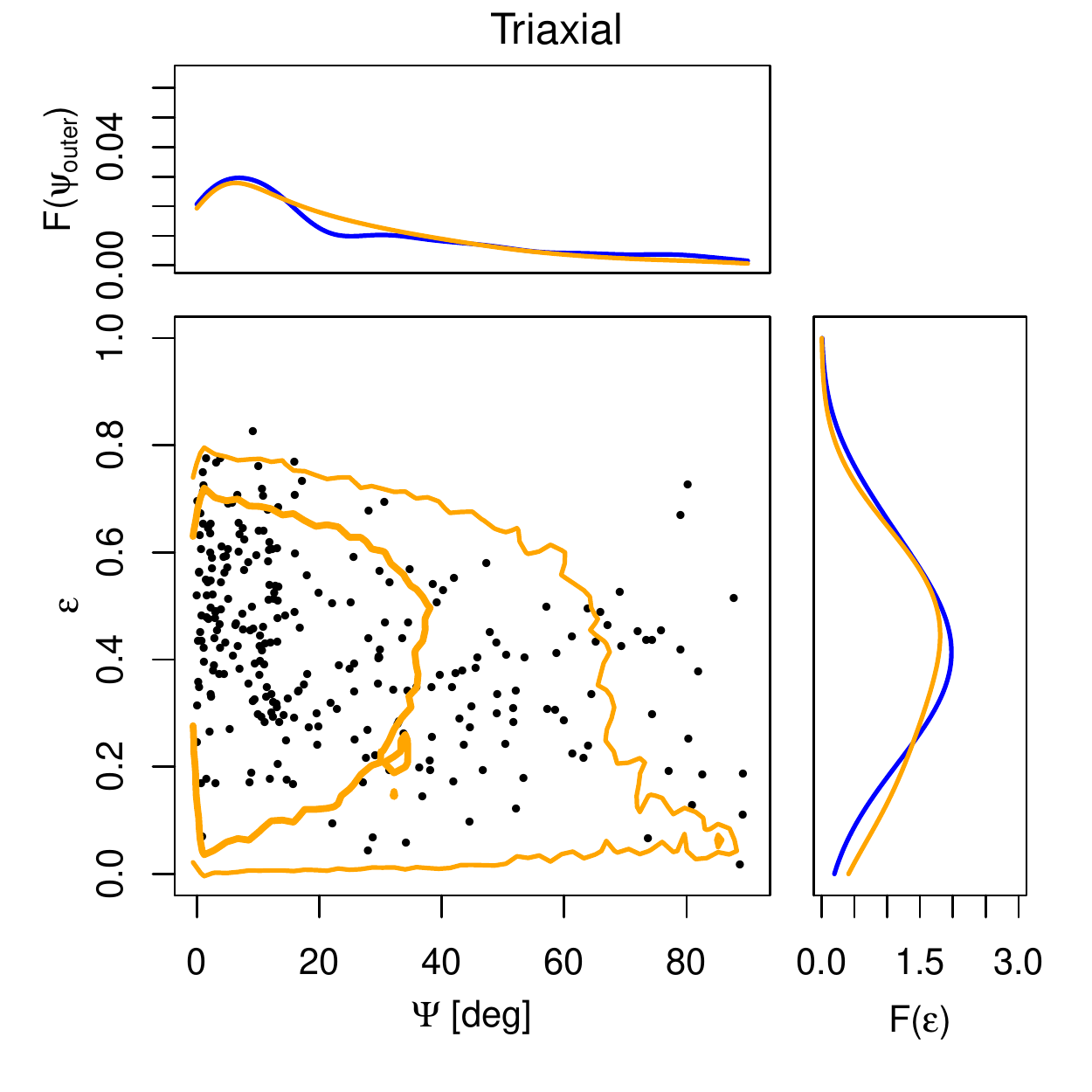}
\caption{Fitting for the intrinsic shape ($p$ and $q$) and $\Psi_{\rm int}$ of galaxy samples separated by intrinsic shape based on $p$ and $q$. Here we show the distributions for samples that have not been cleaned of barred and disturbed galaxies. The distributions of the mock observables $\Psi$ and $\epsilon$ are shown in the top and right panels with the input (blue) and fitted (orange) smoothed normalised distributions $F(\Psi)$ and $F(\epsilon)$. Distributions shown in orange with thick and thin lines represent the 68 and 95\%\ probability intervals, respectively.}\label{fig:shape_recovery}
\end{center}
\end{figure*}

In Figure \ref{fig:shape_recovery} we Illustrate the fitting procedure for the four subsamples of Illustris galaxies presented in the top left panel of Figure \ref{fig:pq_cbars}: spherical, oblate, prolate, and triaxial galaxies. The top and right panels for each shape subsample show the input (blue) and fitted (orange) distributions of $\Psi$ and $\epsilon$, respectively, which have been smoothed and normalised. The central panel for each shape subsample show individual mock observations of $\epsilon$ and $\Psi$ included in the fits. Although we produce 50 mock obervations of each galaxy, to better match observational studies we randomly select one random viewpoint for each galaxy. In this way, each galaxy is represented only once. Thick and thin contours in each of the central panels represent the 68 and 95\% probability intervals of our fits, respectively. Fits are performed following the procedure outlined in Section \ref{section:shapefitting} using both new and old assumptions, that is Equation \ref{eq:thetaint} (old) and fitting for $\Psi_{\rm int}$ explicitly (new). 

\begin{figure*}
	\includegraphics[width=2\columnwidth]{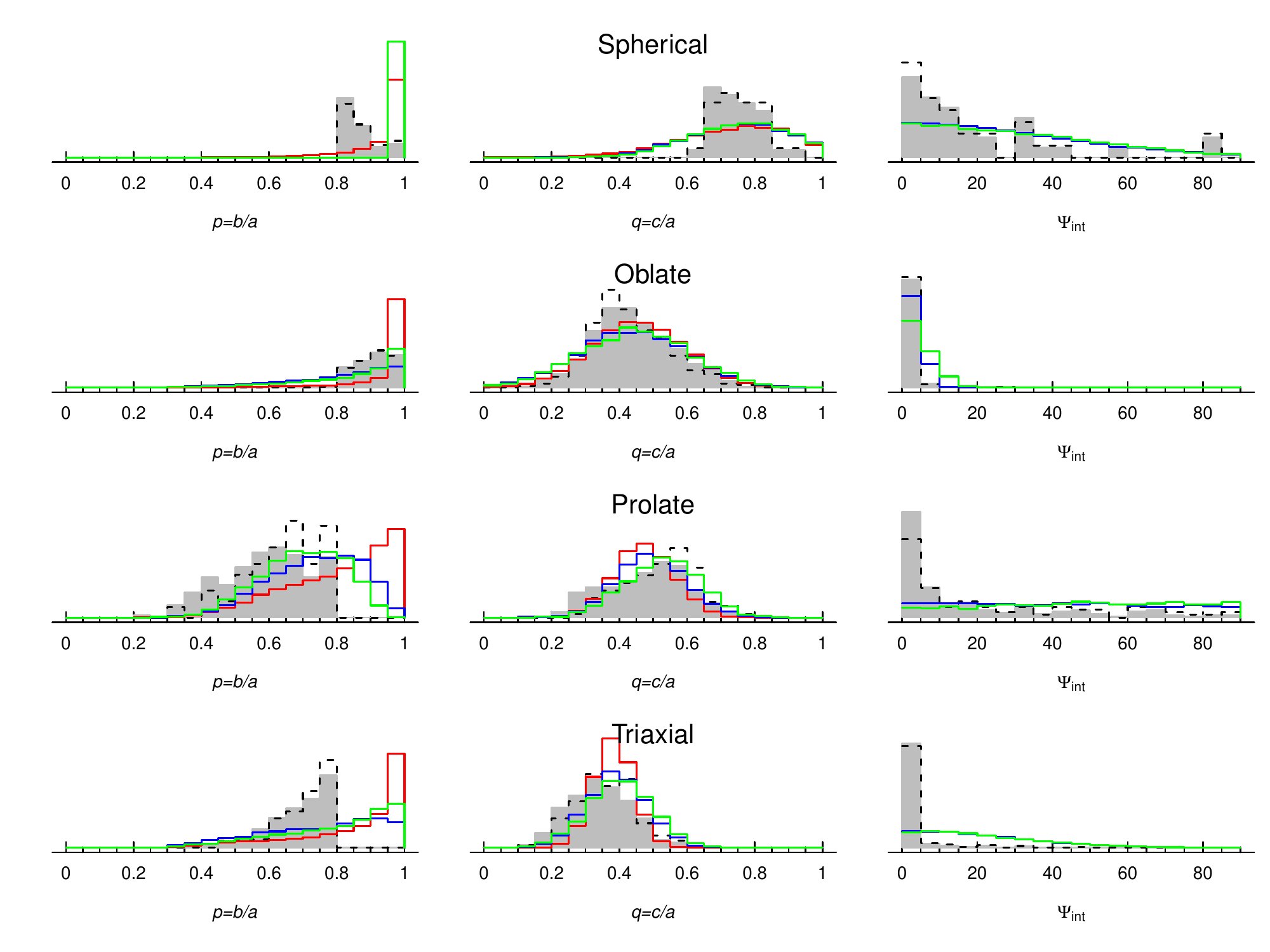}
    \caption{Histograms showing the input (filled grey) cleaned (dashed) intrinsic shape (left and centre) / kinematic misalignment (right) distributions, and the recovered distributions using Equations 16 (red) or fitting $\Psi_{\rm int}$ as a free parameter (blue) and without barred/disturbed galaxies (green).  }\label{fig:pq_recovery}
\end{figure*}

The results of our fitting for each shape subsample are summarised in Table \ref{tab:table1}, where we have also included fit values for fits employing the previous assumption given by Equation \ref{eq:thetaint}. The goodness of each fit is quantified by the values $A^{2}$ (see Section \ref{section:shapefitting}), which can be used to roughly assess whether or not our new assumption has improved the match to the observed distributions of $\Psi$ and $\epsilon$. Also, we note that the results in Table \ref{tab:table1} are for a single realisation of our fitting (i.e. one set of randomly selected viewpoints). We have tested the effect of selecting a different ensemble of random viewpoints, however, finding that the results are qualitatively the same each time, thus our conclusions will remain the same.

For oblate and triaxial galaxies we do not find a large difference in $A^{2}$ between the old and new assumption with oblate changing from 0.0005 to 0.0010 with similar results for triaxial galaxies  (0.0010 vs 0.0011). Spherical galaxies have a poorer fit with $A^{2}$ increasing from 0.0037 to 0.0046, though it should be noted that the spherical sample includes only 26 galaxies and is not well fit in general. Prolate galaxies exhibit the largest improvement between the old and new assumption with $A^{2}$ dropping from 0.0034 to 0.0011. We also show the fit values for our new assumption where galaxies flagged as either disturbed or barred have been removed from the sample. We find that for triaxial and oblate galaxies $A^{2}$ is marginally improved after cleaning the sample while spherical and prolate samples both have a slightly worse fit. We attribute the higher $A^{2}$ value for cleaned spherical and prolate samples, however, to the significant reduction in the sample size, which has effectively increased the noise of the $\Psi$-$\epsilon$ distributions.

Next we take a closer look at the input and fit distributions of $\epsilon$, $F(\epsilon)$, in Figure \ref{fig:shape_recovery}. In general we are able to recover the $\epsilon$ distribution reasonably well. Oblate, Prolate and spherical galaxies are particularly well fit in $\epsilon$ while the fitted $F(\epsilon)$ for triaxial galaxies is more skewed, peaking at a slightly higher value.

Focusing now on input and fit distributions of $\Psi$, $F(\Psi$), we find in most cases our fits are less representative of the underlying distribution when compared to fits to $F(\epsilon)$. The observed distributions for spherical, prolate, and triaxial subsamples have multiple peaks with the global maximum found at $\Psi \sim 0.0^{\circ}$. This is due to small sample statistics and leads to higher values of $A^2$. The underlying distribution for oblate galaxies has a stronger, narrower peak at $\Psi \sim 0.0^{\circ}$. Considering the fit distributions, oblate galaxies exhibit the best recovery of the underlying distribution of $\Psi$. The next best fit is for triaxial galaxies, with $F(\Psi)$ roughly matching the peak location at $\Psi \sim 0.0^{\circ}$ while not matching the width with the fit being significatly broader. For spherical galaxies, aside from the peak at $\Psi = 0^{\circ}$, the value of $\Psi$ is essentially random. In this case the fit to $\Psi$ roughly matches the peak position of the underlying data and attempts to account for the remaining data using a very broad distribution. The results for prolate galaxies are similar however the underlying distribution is less peaked at $\Psi = 0^{\circ}$ and is, overall, more flat. As such, the output $\sigma_{\Psi_{\rm int}}$ is extremely broad at 81.0$^{\circ}$, roughly twice that of the next highest value found for spherical galaxies of 40.6$^{\circ}$. 



We show in Figure \ref{fig:pq_recovery} the input and fitted distributions of $p$ and $q$ for each of our shape subsamples. The original and cleaned input distributions are shown as the shaded grey region and dashed histogram, respectively, while the fits for the old assumption, the new assumption, and the new assumption with disturbed and barred galaxies removed are shown with red, blue, and green lines, respectively. The cleaned and original input distributions are usually similar. This underscores the point from above that removing disturbed and barred galaxies from most of our samples does not seem to improve our results appreciably. Such galaxies can not always be identified from observations, depending on their exact orientation, and may be included in observed samples. Thus, failing to remove such objects from an observational dataset will not significantly affect the ability of this technique to recover the shape of the galaxy sample. 

How well do we recover the shapes of each galaxy subsample? Considering the recovery of $p$ for spherical and oblate galaxies, we find that both the old and new assumptions result in a distribution of $p$ that is skewed towards $p=1$, though less so for the new assumption for oblate galaxies. The underlying distributions, however, are more widely distributed between 0.8 and 1, the boundaries for our definitions for spherical and oblate galaxies. Prolate and triaxial galaxies are defined as having $p<0.8$. Fits using the new assumption better capture this constraint with the $p$ distributions moving away from $p=1$ for prolate galaxies, however a preference for $p=1$ for triaxial fits remains. Using the old assumption we again find fit $p$ values for triaxial galaxies are skewed towards $p=1$ and the prolate fit includes a significant representation of $p>0.8$ galaxies inconsistent with the definition for these subsamples. For the old assumption, this is due to the presence of $\Psi_{\rm int}\sim0.0^{\circ}$ galaxies in these subsamples, forcing $p\sim1$ (thus $T\sim0$) for kinematically aligned galaxies. For the new assumption this could be due to the fact that we assume $p$ is lognormally distributed though this is clearly not the case looking at the underlying distribution.

The ability of our method to recover $q$ also varies between galaxy subsamples. In general the resulting distributions are quite similar for all three fits. We find that oblate galaxies exhibit the best agreement between the underlying and fit distributions for $q$ while spherical galaxies exhibit the worst agreement. The poor fit for spherical galaxies may be due to the small sample size or the fact that the underlying $q$ distribution is not well described by a Gaussian. For prolate galaxies the fits for the old assumption and the new assumption without barred/disturbed galaxies removed agree well but underestimate the median value of $q$. After removing barred/disturbed galaxies, the output distribution is better matched to the underlying distribution likely due to effects of galactic bars on measurements of $p$, $q$, and $\epsilon$. For triaxial galaxies the median of all three fits is in rough agreement though slightly higher than the underlying distribution. Fits using the new assumption result in a slightly broader distribution that is better matched to the true $q$ distribution. Overall, the comparison of our fit distributions with the underlying distributions indicates that our choice of a normal distribution may not be ideal.

How well can we recover $\Psi_{\rm int}$? We also note that the $\Psi_{\rm int}$ distributions are reasonably recovered for oblate and spherical galaxies, but fail miserably for prolate and triaxial systems. Similar to the recovery of $q$ for our subsamples, this indicates that an assumption of a Gaussian distribution is not well suited to this problem or that there isn't enough information in the data to constrain all 6 free parameters.

Overall, it seems that our new assumption can provide a slight improvement in recovering the shape of our galaxy subsamples, though not in all cases. In particular the recovery of $p$ for prolate galaxies is in better agreement with the definitions of this subsample. This is primarily due to the allowance for $\Psi_{\rm int}\sim0.0^{\circ}$ for a wider range of galaxy shape. There are a number of discrepancies between the underlying and recovered distributions of $p$ and $q$, however, that put the viability of our method into question. The key issue, highlighted in Section \ref{section:what_shapes_galaxies} is the lack of a clean relationship between shape and $\Psi_{\rm int}$. Figure \ref{fig:Psi_T} suggests that $\Psi_{\rm int}$ is either 0.0$^{\circ}$ or randomly distributed. This is consistent with the analysis presented in Section \ref{section:mergerhist}, which shows that kinematic offsets in galaxies can be attributed to mergers and that the final $\Psi_{\rm int}$ may also relate to the random orientation of the merger orbit.

We reiterate that a straightforward improvement to the method tested here is to leave $\Psi_{\rm int}$ as a free parameter \citep[as was done in][]{Li18a} given that we show in Figures \ref{fig:Psi_T} that there is no simple relationship between $\Psi_{\rm int}$ and intrinsic shape. Other possible avenues for shape recovery may include supervised machine learning using full stellar luminosity and stellar kinematics maps (rather than single quantities derived from these, i.e. $\Psi$ and $\epsilon$) or Bayesian methods for recovering posterior distributions of $p$ and $q$ from other metadata (e.g. mass, gas fraction, colour, etc.). We discuss these possibilities further in Section \ref{section:future}. 

\section{Discussion}\label{section:discussion}

\subsection{Simulated vs Observed Galaxy Shapes}\label{section:sim_v_obs}

\begin{figure*}
	\includegraphics[width=1.8\columnwidth]{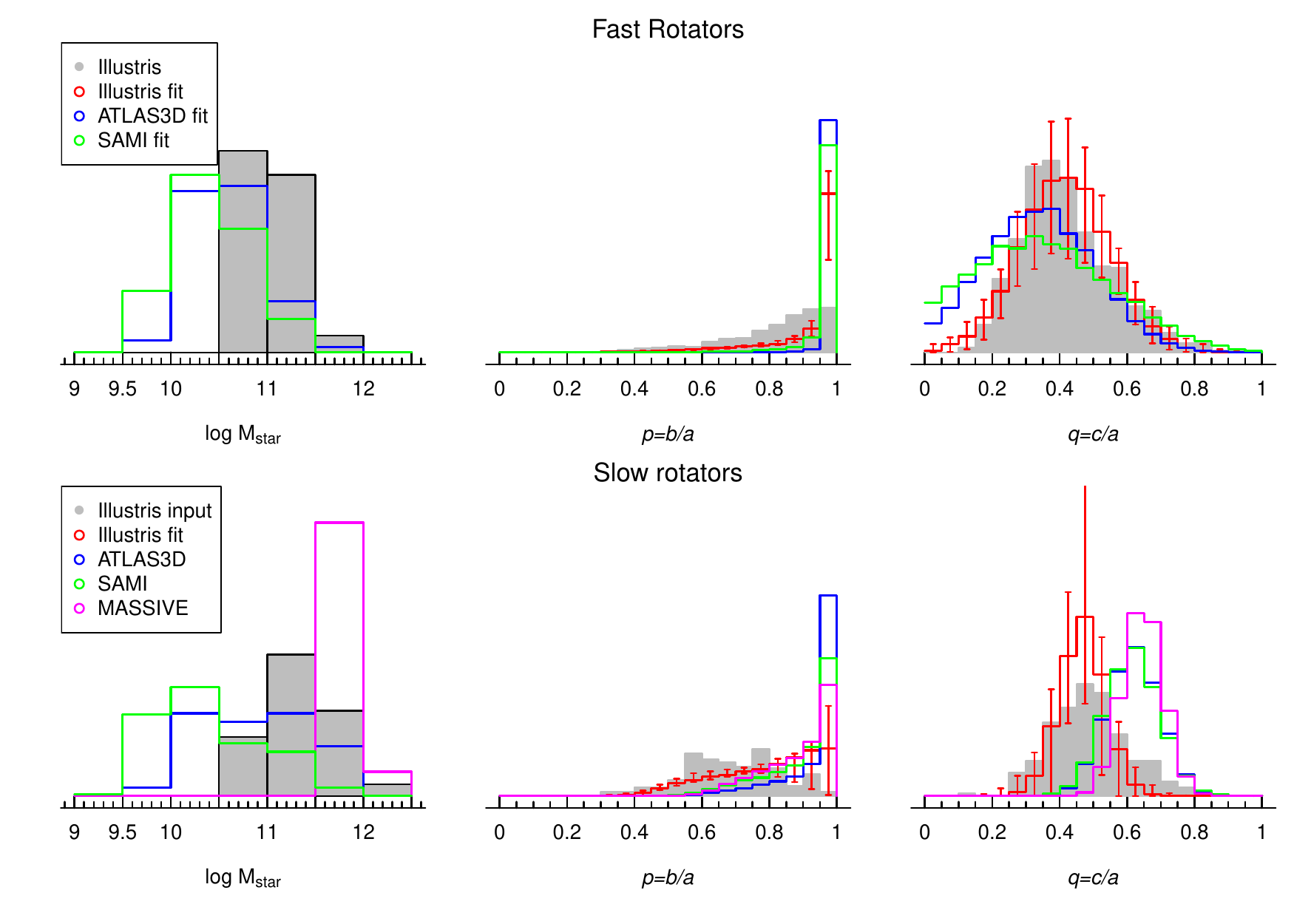}
    \caption{Comparing to previous literature for fast and slow rotators. Despite clear differences in stellar masse distributions (left) probed by the various surveys, fitted distributions of $p$ (middle) and $q$ (right) for fast (top row) and slow (bottom row) samples in the literature tend to agree better with each other than with the simulations. Red errorbars are not uncertainties, instead they represent the range of distributions from 25 random orientations with the red histograms showing the median solution.}\label{fig:obscomp}
\end{figure*}

Having tested the ability of our method to recover the shapes of Illustris galaxies from projected quantities, we compare these results with recovered distributions from observations of real galaxies. There are two important caveats inherent in the comparison between Illustris galaxy shapes and the observational works presented. First, we compare to studies from SAMI, ATLAS3D, and MASSIVE, each of which are based on specific, but different, volume, stellar mass, and morphological selections (e.g. ATLAS3D and MASSIVE focus exclusively on early type galaxies). Imposing comparable selection biases to Illustris galaxies is beyond the scope of this work. Second, simulated galaxies from Illustris may not have an intrinsic shape distribution matching the real universe. Such differences in the underlying shape distributions of Illustris and real galaxies will result in differences in the ensemble of projected luminosity and kinematics maps when compared with observational studies. Thus, we caution against over-interpretation of the comparative results presented here. 

\subsubsection{Recovered Shape Distributions}

We show in Figure \ref{fig:obscomp} a comparison between the known $p$ and $q$ distributions from Illustris in grey and fitted distributions to both mock observational data presented in work and true observational efforts from the SAMI \citep{Foster17}, ATLAS3D \citep{Weijmans14}, and MASSIVE \citep{Ene18} teams. We have split between slow and fast rotators on the top and bottom rows respectively noting that the MASSIVE survey did not include enough fast rotators to perform the fit. For any single Illustris galaxy, the classification as slow or fast rotator may depend on viewing angle. Thus, for the simulated sample we randomly select a single viewpoint for each galaxy and use the projected values for this viewpoint for both classification and shape recovery fitting. We also note that here we have performed our shape recovery analysis with the old assumption of Equation \ref{eq:thetaint} (i.e. $T=(1-p^{2})/(1-q^{2})$ correlates with $\Psi_{\rm int}$). Thus, results from observational studies and from this work shown in Figure \ref{fig:obscomp} are obtained using the same procedure.

We also show in the left column of Figure \ref{fig:obscomp} a comparison of the stellar mass distributions of the observed samples with our Illustris sample. In the case of SAMI and ATLAS3D, the stellar mass of both fast and slow rotators probe a slightly lower mass regime when compared to the Illustris galaxies explored here. MASSIVE galaxies, on the other hand, probe exclusively the high mass end of our Illustris sample for slow rotators. We reiterate that reproducing the various selection effects of these observational works for our Illustris sample is beyond the scope of this work, and thus we simply leave this as a caveat to the comparison presented here.

First we consider the recovered distributions of $p$ for fast rotators shown in the top left panel of Figure \ref{fig:obscomp}. For fast rotators, the fit distributions for both observed and simulated are strongly skewed towards $p=1$. As we have shown in Figure \ref{fig:Psi_T}, this is most likely a result of the assumed relationship between $\Psi_{\rm int}$ and intrinsic shape from Equation \ref{eq:thetaint} as per \citet{Weijmans14}. Under this assumption, $\Psi_{\rm int} = 0^{\circ}$ only occurs where $T=1$, which is equivalent to $p=1$. This means that the assumption of Equation \ref{eq:thetaint} combined with the fact that the vast majority of fast rotators have $\Psi \simeq 0^{\circ}$ results in this skewed distribution. Thus, the recovered $p$ distribution for fast rotators is more related to the assumption rather than the underlying distribution. For this reason the $p$ distribution for fast rotators gives little insight into the similarity of underlying shape distributions between observed and simulated galaxy samples.

Compared to the fits for fast rotators, we find a slightly larger discrepancy between the fit distribution of $p$ for simulations and observations in slow rotators. For Illustris we perform our fits 25 times, each time selecting galaxies at a new, randomly selected ensemble of viewpoints to classify our FR and SR samples. We plot the median distribution in red and show the lower and upper limits in each bin with the red errorbar noting that this is not an ``error", but simply illustrates the range of possible solutions stemming from systematics associated with randomly selected viewpoints. We show a single realisation of the underlying distribution in grey, noting that this distribution varies little between iterations. Fit distributions for observed samples have a strong peak at $p=1$, similar to fast rotators, but taper off more gradually towards lower $p$ compared to distributions for fast rotators. This suggests that even in samples of observed slow rotators we find a preference for kinematic alignment. 

Given that the same fitting procedure has been followed for all fits shown in Figure \ref{fig:obscomp}, it is likely there is either a difference between the distribution of $\Psi$ in the projected Illustris sample or a difference between the underlying distributions of intrinsic shape and/or $\Psi_{\rm int}$. The former case could occur if the individual realisations of our Illustris slow and fast rotator samples used for the fit in Figure \ref{fig:obscomp}, where each galaxy is sampled at a single viewpoint and classified accordingly, exhibited an overabundance of projections with large $\Psi$. This would result in fit distributions skewing towards lower $p$ as here we assume the relationship between $\Psi_{\rm int}$ and intrinsic shape given in Equation \ref{eq:thetaint}, which implicitly presumes $\Psi > 0$ necessitates $p \ll 1$. Given the underlying distribution here is well matched to the fit, and that such a large over-representation is rare for our single viewpoint sampling (we estimate this to occur in $\sim$2\% of all samples), we consider the latter possibility more likely (though not certain): the underlying distributions of intrinsic shape and/or $\Psi_{\rm int}$ are different between Illustris and the observed universe. If true, differences in underlying shape may also stem from differences in sample selection and the related difference in the mass distribution of observed and simulated samples. Thus, we refrain from making and strong conclusion regarding differences in $p$ between Illustris and real galaxies. 

We have shown in Section \ref{section:altass} that the recovery of $q$ is generally more reliable than the recovery of $p$, most likely because it is less dependent on the assumed relationship between intrinsic shape and $\Psi_{\rm int}$. We again find this to be the case for our Illustris fast and slow rotator samples, with the fits providing a reasonable recovery of both the mean and spread in the underlying $q$ distributions. There is, however, a significant difference between the distributions of $q$ for Illustris galaxies when compared to the fits for observed samples. Illustris fast rotators have a slightly higher mean $q$ value compared to observed samples and $q < 0.2$ galaxies are not present in the simulations. Although such extremely flat galaxies are relatively rare in observations, they do exist \citep[e.g.][]{Goad81,Karachentsev99,Rodriguez13}. The lack of similar galaxies in cosmological hydrodynamical simulations may result from resolution limits, prescriptions for gas cooling, or a combination of the two \citep{Stinson13,Marinacci14}, but is a known discrepancy between simulated and observed galaxies. The mean $q$ for Illustris slow rotators is also found to disagree with fits for observed samples, this time being significantly lower. As we have stated, in general $q$ in Illustris is well recovered, thus we should not expect the recovery for observed samples to be significantly worse given the same method has been used for all fits in Figure \ref{fig:obscomp}. Instead, differences in fits for $q$ between Illustris and observations is further evidence that the underlying shape distribution of Illustris galaxies may not be representative of the real universe.


\subsubsection{Underlying Shape Distributions}

Given the comparisons presented above, then, is the distribution of intrinsic shapes in Illustris found to be representative of real galaxy shapes? One would expect that, if the underlying shape distributions of Illustris and the real universe were the same, the fitted distributions in both $p$ and $q$ would be roughly the same as well given that the fits are produced in an identical fashion using different datasets. Thus, it is likely that Illustris galaxies do not have a shape distribution representative of real galaxies. We must stress, however, that this conclusion is based on the comparison between the fit distributions with no direct knowledge of the underlying shape distributions of real galaxy samples. Furthermore, we reiterate that differences in sample selection and mass distributions of observed versus Illustris galaxies may also affect the underlying shape content of various samples.

Related to this, one may ask if fit distributions to observed galaxies are actually reliable? This question is difficult to answer conclusively based on our analysis due again to the fact that the underlying 3D shape distribution of real galaxy samples is not known. It is clear that the fit distribution of $p$ for Illustris galaxies is more dependent on the assumption of Equation \ref{eq:thetaint} and the presence or absence of $\Psi \neq 0$ among input projections (particularly at higher $\epsilon$). Thus, while we cannot rule out the possibility that the underlying distribution of $p$ for observed galaxies does match the fits presented in \citet{Weijmans14}, \citet{Foster17}, and \citet{Ene18}, we can confidently state that a broader underlying distribution would not be recovered assuming Equation \ref{eq:thetaint}. Considering the recovery of $q$, we find that for the majority of our Illustris subsamples we are able to reliably recover the underlying $q$ distribution. Thus, our analysis suggests that $q$ is more likely to be reliably recovered for a given sample than $p$, though we can not definitively state that observed $q$ distributions are representative of the intrinsic distributions.

\begin{figure}
	\includegraphics[width=\columnwidth]{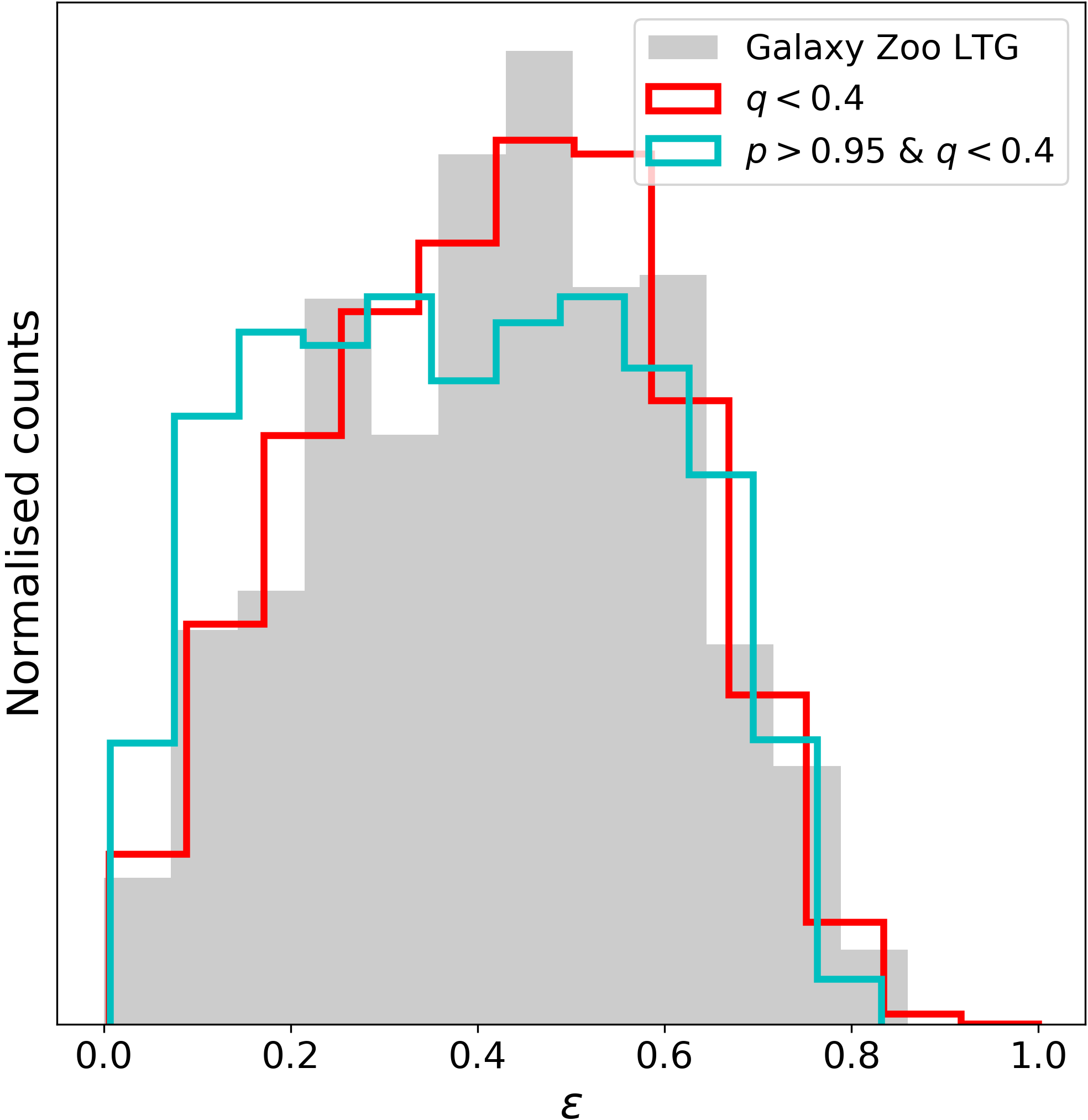}
    \caption{Comparing distributions of observed axis ratios of SDSS late-type galaxies from 2MASS data (grey filled histogram) with subsamples of mock observations from Illustris. The observational data includes all galaxies for which more than 2/3 of all classifications from the Galaxy Zoo project are late-type (both face-on and edge-on included) while the two Illustris subsamples are selected based on 3D axis ratios. Extremely-round, oblate Illustris galaxies with $p>0.95$ have a flat distribution in axis ratio between $\sim$0.3 and $\sim$0.95 while observed late-type galaxies peak between 0.4 and 0.5 and fall off towards higher axis ratio. This suggests the underlying distribution of $p$ for real disk galaxies include a significant number of galaxies with $p\neq 0$. We also show mock axis ratios for flattened Illustris galaxies with $q<0.4$, which exhibit a range in $p$, are better matched to the observed axis ratios.}\label{fig:eps_hist}
\end{figure}

A related issue relevant to a number of astrophysical questions, particularly those relying on accurate inclination corrections, is: are face-on galaxies intrinsically round (i.e. $p=1$)? In Figure \ref{fig:eps_hist} we show the observe $\epsilon$ of flat ($q < 0.4$, red) Illustris galaxies, as well as the subset of flat and extremely round ($q < 0.4$ and $p > 0.95$, cyan) galaxies. The former distribution peaks around  0.6 while the latter is flat between 0.2 and 0.8. We show the underling distribution of $\epsilon$ for late-type galaxies identified by the Galaxy Zoo project \citep{Lintott08,Lintott11} averaged over J, H, and K imaging from the Two Micron All-Sky Survey \citep[2MASS,][]{Skrutskie06}, finding it better matched to the sample including all $p$ values. This suggests galaxies are not intrinsically round, however, we do not draw any strong conclusions from this for two reasons: first, proper comparison matching various sample selection criteria is required, and second, we have not accounted for biases associated with morphological classification of Sloan Digital Sky Survey \citep[SDSS,][]{York00} images. While interesting, a rigorous comparison of observed and simulated $\epsilon$ is beyond the scope of this work. We comment however that \citet{Rodriguez13} have analysed the projected shapes of SDSS galaxies inferring that large numbers of galaxies of all morphological types exhibit $p \neq 1$. 

Finally, we comment on the application of the new assumption for the $\Psi_{\rm int}$ distribution  presented in Section \ref{section:altass} to observed galaxy samples.
Again, the key here is that $\Psi_{\rm int} = 0.0$ is allowed at a wider range of shapes when compared to Equation \ref{eq:thetaint} with the understanding that $\Psi_{\rm int}$ depends on specifics of a galaxy's history (mergers in particular, see Section \ref{section:mergerhist}) rather than it's shape. We find that for Illustris galaxies the assumption of a normally distributed $\Psi_{\rm int}$ provides a fit to the underlying shape distribution that is either equivalent or better than the previous assumption (i.e. Equation \ref{eq:thetaint}), though still not exact. We find a slight improvement in the recovery of $p$ for prolate and triaxial galaxies as shown in Figure \ref{fig:pq_recovery} and a slight reduction in the goodness of fit value, $A^{2}$. Given the random behavior of $\Psi_{\rm int}$ for all but oblate galaxies, we conclude that leaving $\Psi_{\rm int}$ as a free parameter \citep[e.g.][]{Li18a} is preferable.

\subsection{Future Directions for Intrinsic Shapes Studies}\label{section:future}

In this work we have explored the validity of the assumed relationship between $T$ and $\Psi_{\rm int}$ utilised in studies that attempt to recover 3D galaxy shapes for samples of galaxies observed with IFS. We have shown in Section \ref{section:what_shapes_galaxies} that this assumption is not valid, at least for galaxies in the Illustris-1 simulation. What this importantly shows is that the assumption that static St\"{a}ckel potentials accurately describe galaxies \citep[e.g.][]{Franx91} does not hold. Indeed, the potential of any galaxy in a given hydrodynamical simulation can vary significantly as a function of time. Given the large range of stable configurations in Illustris where Equation \ref{eq:thetaint} is not valid, we expect a similarly wide range of stable configurations to be present in samples of real galaxies.

In Section \ref{section:altass} we have tried fitting $\Psi_{\rm int}$ explicitly to avoid the pitfall that galaxies with $p\neq 1$ must correspond to large $\Psi_{\rm int}$. Though this new assumption provides a slight improvement in the recovery of Illustris galaxy shapes, striking discrepancies between the underlying and recovered shape distributions remain (see Figures \ref{fig:pq_recovery} and \ref{fig:obscomp}). These results beg the question: can we identify a viable way forward for future galaxy shape recovery efforts?

Beyond simultaneously fitting for $\Psi_{\rm int}$, another possible avenue is the use of Bayesian statistics, a technique that is growing rapidly in popularity within the astronomy community \citep[e.g.][among many others]{Feroz07,Cameron12,Jennings16,Thomas18}. Bayesian techniques would allow us to calculate posterior probability distributions and quantify degeneracies and uncertainties for $p$, $q$ and $\Psi_{\rm int}$ and use priors inferred from a larger set of galaxy properties than $\epsilon$ and $\Psi$ alone. We have shown in Figure \ref{fig:pq_cbars}, for example, that $q$ is well correlated with galaxy specific angular momentum. Other possible galaxy properties include star formation rate, stellar mass,  galaxy colour, S\'{e}rsic index, or any other property that shows some dependence on galaxy morphology. We note that \citet{Rivi18} present an example of a similar analysis employing a Hamiltonian Monte Carlo Bayesian analysis to estimate the intrinsic ellipticities of star-forming galaxies from radio continuum survey visibility data. Using prior distributions for ellipticity and galaxy scale length, the authors are able to accurately recover the intrinsic ellipticities in simulated data. Thus, a precedent for Bayesian shape recovery already exists in the literature. 

Another possible direction for future shape recovery studies is using machine learning techniques, which can better parse datasets with very high dimensionality \citep[though with measured caution, e.g.][]{Gao17}. In this way, we may be able to develop new shape recovery algorithms using full stellar luminosity and kinematics maps (both velocity and velocity dispersion) rather than distilling the complexity contained in these datasets into a single pair of observables per galaxy ($\epsilon$ and $\Psi$) as is currently done. We note that machine learning techniques are routinely applied to astronomical imaging data to solve a variety of problems including morphological classifications \citep[e.g.][]{DominguezSanchez18,Lukic18} and selections of lensed galaxies \citep[e.g.][]{Jacobs17,Pourrahmani18}. Recently, \citet{Wu18} employed convolutional neural networks to predict the metallicities of galaxies from their three-colour images alone with highly accurate recovery. If a similar approach could be used to recover galaxy shapes, this would be an ideal application for constraining the shapes of source galaxies in cosmological weak lensing studies where available source properties beyond colour and photometric redshift are often not available. Another promising area is that of generative adversarial networks \citep{Goodfellow14}, which have been applied, for example, by \citet{Zhang18} in the non-astronomy context of generating 3D hairstyle structure from a single 2D image. Although the details of this work are significantly different than that of 3D galaxy shapes, the problem is conceptually similar to that presented in this work. The key caveat here is that these methods will be ``tuned'' to the simulations used for the training set. As we have mentioned in Section \ref{section:sim_v_obs}, the underlying shapes of Illustris galaxies are likely different from galaxies in the real universe (e.g. the previously mentioned lack of extremely flattened galaxies in Illustris).

In future work we will apply both Bayesian and machine learning techniques to mock images produced from Illustris and other cosmological hydrodynamical simulations with the aim of providing novel shape recovery methods. These methods will again be tested using mock observations of galaxies from cosmological hydrodynamical simulations, however. Thus, a better understanding of both the differences between true galaxy shapes and those found in cosmological simulations as well as how our shape recovery methods depend on these differences will be required before such techniques can be reliably applied to observed datasets. 

\section{Conclusions}\label{section:conclusions}

In this work we have examined the shapes of Illustris galaxies in three dimensions with the aim of testing methods of recovering galaxy shape from projected IFS observations \citep[e.g.][]{Weijmans14,Foster17,Ene18}. We began by exploring the distributions of galaxy shapes at the $z=0$ snapshot of the Illustris-1 simulation, the highest resolution run from the Illustris project \citep{Vogelsberger14a,Vogelsberger14b,Genel14}. In particular, we placed a strong emphasis on the relationship between galaxy shape and kinematic misalignment as quantified by $\Psi_{\rm int}$, the angle between the two vectors describing a galaxy's angular momentum and ellipsoidal minor axis. As an example, for an oblate (disk) galaxy with normal rotation within the disk plane, one can expect $\Psi_{\rm int} \simeq 0^{\circ}$. Our key findings here are:
\begin{itemize}
	\item We find the strongest correlation is between specific angular momentum, $J/M_{*}$, and $q$ where more rapidly rotating galaxies at fixed mass are also more flattened
    \item 67\% of kinematically offset galaxies (those with $\Psi_{\rm int} > 30^{\circ}$) are prolate in shape, though $\Psi_{\rm int} \simeq 0^{\circ}$ prolate galaxies are also found in Illustris-1
    \item Similar to \citet{Li18b}, we find that the transition to prolate shape is related to a past, major-merger, and this merger is also the origin of strong kinematic offsets in $\Psi_{\rm int} > 75^{\circ}$ galaxies
    \item For $\Psi_{\rm int} < 5^{\circ}$, prolate galaxies, a major merger is also responsible for a transition to a prolate shape. One typical difference between kinematically aligned and misaligned prolate galaxies is that aligned prolate galaxies harbour more neutral gas prior to the merger, though merger geometry likely also plays a significant role
    \item The assumed relationship between $\Psi_{\rm int}$ and galaxy shape (Equation \ref{eq:thetaint}) employed in previous work aimed at recovering galaxy shape from IFS observations does not hold for Illustris-1 galaxies (see Figure \ref{fig:Psi_T})
\end{itemize}

After characterising the shapes and kinematic offsets of galaxies from the Illustris simulation, we next tested one of the methods for shape recovery commonly employed in the literature. This was done by producing mock $r$-band luminosity and stellar kinematics maps for Illustris galaxies with random orientations, simulating observations from IFS surveys \citep[e.g. SAMI,][]{Croom12,Scott18}. As the previous relationship between galaxy shape and $\Psi_{\rm int}$ was shown to be invalid in Illustris, we also performed fits to the same dataset with $\Psi_{\rm int}$ left as a free parameter in an attempt to improve the shape recovery for simulated galaxies. The results of our analysis of shape recovery from projected mock images are summarised as: 
\begin{itemize}
	\item The assumed relationship between $\Psi_{\rm int}$ and galaxy shape (Equation \ref{eq:thetaint}) biases fits to $p\simeq$1 due to the majority of galaxies having $\Psi_{\rm int}\simeq 0^{\circ}$
    \item We recommend fitting for $\Psi_{\rm int}$ explicitly to allow for a larger range in $p$ at fixed $\Psi_{\rm int}\simeq 0^{\circ}$, which gives a slight improvement in $p$ recovery
    \item The mean fit value of $q$ is similar under both assumptions and both provide a reasonable match to the underlying mean
    \item For prolate and triaxial galaxies, the fit $q$ distribution is broadened under the new assumption resulting in a better agreement with the underlying distribution
\end{itemize}
Regardless of the marginal improvements in shape recovery under an alternative assumption for $\Psi_{\rm int}$, the overall performance of the method is not ideal. This is particularly true for the recovery of $p$. Understanding the distribution of $p$ for galaxy samples is important for science cases requiring accurate inclination corrections as these often rely on the assumption that galaxies are perfectly round when viewed face-on (i.e. $p=1$). 
Given the random behaviour of $\Psi_{\rm int}$ for prolate galaxies, we suggest that leaving $\Psi_{\rm int}$ as a free parameter while fitting rather than coupling it to galaxy shape may provide a slightly more reliable intrinsic shape recovery \citep[e.g.][]{Li18a}. Other possible avenues include reconfiguring our fitting code to include a Bayesian analysis or employing generative adversarial neural networks trained on simulated galaxies. We plan on testing each of these methods in future work.

\section*{Acknowledgements}

This research was conducted by the Australian Research Council Centre of Excellence for All Sky Astrophysics in 3 Dimensions (ASTRO 3D), through project number CE170100013. This publication makes use of data products from the Two Micron All Sky Survey, which is a joint project of the University of Massachusetts and the Infrared Processing and Analysis Center/California Institute of Technology, funded by the National Aeronautics and Space Administration and the National Science Foundation. This research made use of Astropy, a community-developed core Python package for Astronomy \citep{Astropy13}; matplotlib, a Python library for publication quality graphics \citep{Hunter07}; SciPy \citep{Jones01}. 





\bibliographystyle{mnras}
\bibliography{biblio} 


\bsp	
\label{lastpage}
\end{document}